\documentclass[11pt]{article}
\usepackage[top=1in, bottom=1in, left=1in, right=1in]{geometry}
\usepackage{amsmath}
\usepackage{amsbsy}
\usepackage{amscd}
\usepackage{amsopn}
\usepackage{amstext}
\usepackage{amsxtra}
\usepackage{mathrsfs}
\usepackage{graphicx}
\usepackage{epstopdf}
\usepackage{epsfig}
\usepackage{authblk}
\usepackage{cite}
\usepackage{url}
\usepackage{hyperref}
\usepackage{amsmath}
\usepackage{amsbsy}
\usepackage{amscd}
\usepackage{amsopn}
\usepackage{amstext}
\usepackage{amsxtra}
\usepackage{color}
\usepackage{lscape}
\usepackage{authblk}

\begin{document}

\title{Branching process descriptions of information cascades on Twitter}

\author{James P Gleeson$^{1,2,3}$,  Tomokatsu Onaga$^4$, Peter Fennell$^{1,5}$, James Cotter$^1$, Raymond Burke$^1$, David J.~P.~O'Sullivan$^1$}
\affil{$^1$ MACSI, Department of Mathematics and Statistics, University of Limerick, Ireland.\\
$^2$ Insight Centre for Data Analytics, University of Limerick, Ireland.\\
$^3$ Confirm Centre for Smart Manufacturing, University of Limerick, Ireland.\\
$^4$ The Frontier Research Institute for Interdisciplinary Sciences \& Graduate School of Information Sciences, Tohoku University, Japan.\\
$^5$ USC/ISI, 4676 Admiralty Way, Marina Del Rey, Los Angeles, California 90292, U.S.A. }

\date{13 July 2020}

\maketitle

\begin{abstract}
A detailed analysis of Twitter-based information cascades is performed, and it is demonstrated that branching process hypotheses are approximately satisfied. Using a branching process framework, models of agent-to-agent transmission are compared to conclude that a limited attention model better reproduces the relevant characteristics of the data than the more common independent cascade model. Existing and new analytical results for branching processes are shown to  match well to the important statistical characteristics of the empirical information cascades, thus demonstrating the power of branching process descriptions for understanding social information spreading.

\end{abstract}

\section{Introduction} 

The transmission of information via online social networks is increasingly ubiquitous. The volume of freely-available data  offers unprecedented opportunities for data-driven mathematical modelling of human behaviour. Twitter, for example, is a directed social network wherein users ``follow'' other users in order to receive their broadcast transmissions, called ``tweets''. All tweets are public, making analysis of Twitter data particularly popular among data scientists. Twitter users may retweet messages they receive from the users they follow, and in this way cascades of information may stem from a single tweet event that we call the ``seed''. In the search for mathematical models to describe such structures, branching processes \cite{AthreyaNeybook,Harrisbook} are an appealing option. As stochastic processes, they can potentially capture the wide variability observed in tweeting patterns and human behaviour, while offering a wealth of theoretical results that can be tested against data from online social networks.

Branching processes have already been applied in several studies of Twitter and other online fora. The recent review by Arag\'{o}n et al.~\cite{Aragon17} surveys models of discussion threads, including Twitter reply cascades. Several of the generative models cited in \cite{Aragon17} are based on branching processes, but most find it necessary to modify classical branching processes with some novel features in order to match to data. For example, Nishi et al.~\cite{Nishi16} studied reply cascades in Twitter (as distinct from the retweet cascades that we examine here) that were seeded by celebrities and found that these could not be fitted by classical Galton-Watson processes, so they introduced a modified version of the branching process. On the other hand, Galton-Watson processes (albeit with special seed offspring distributions) were successfully applied to discussion trees from Reddit by Medvedev et al.~\cite{Medvedev18}, and time-dependent continuous-time branching processes were fitted to a viral marketing campaign in \cite{Iribarren09,Iribarren11}. Golub and Jackson \cite{Golub10} reanalysed data from \cite{LibenNowell08} to show that although standard branching processes did not appear to reproduce the features of the email cascades studied in \cite{LibenNowell08}, when selection bias was added---to model the fact that large, viral, chains are more likely to be observed than small chains---then the biased Galton-Watson process fitted quite well.

Although branching processes have been fitted to data to form the basis for simulation and prediction in several studies, the application of analytical results from branching processes theory has been mostly limited to a small selection of features. The most common \cite{Aragon17} is the cascade size, i.e., the accumulated number of tweets or replies to a single seeding post, for which the well-known Galton-Watson result for the expected total number of progeny has been used for prediction, e.g.,  \cite{SeismicZhao15}. However, determining the entire distribution of cascade sizes, not just its mean, is quite feasible \cite{GleesonPRL14,Gleeson16}, as are analytical (or semi-analytical) methods for the calculating the length and depth of cascade trees, as well as other measures. One such measure that we examine in Sec.~\ref{sec4.3} is the ``structural virality'' of a cascade, as introduced by Goel et al.~\cite{Goel15}. Using this and other measures of cascade trees, Goel et al.~performed large-scale numerical simulations of a simple transmission model on networks to fit to data from Twitter. The transmission model of \cite{Goel15} is a discrete-time version of the susceptible-infected-recovered disease-spread model, also known as the ``independent cascade model'' (ICM) \cite{Kempe03}. Other network-based simulation models \cite{Larremore12,Sreenivasan17} use variations of such dynamics (such as susceptible-infected-susceptible disease-spread models \cite{Leskovec07}) to understand the effects of network structure upon spreading.

In this paper we focus on three aspects of branching process models for Twitter retweet cascades, using a reanalysis of two previously-studied datasets \cite{OSullivan17,Hodas14}. First, in Section~\ref{sec2}, we extract the empirical offspring distributions from the tree structures and show that these remain approximately stable across a range of generations. This is a necessary condition for classical branching processes to provide accurate models for detailed features of cascade trees, and the simplicity of this result contrasts with models where explicit time-decay of novelty \cite{WuH07,Yasseri17,GleesonPNAS14} or generation-dependent branching numbers \cite{Dobson12} are required.

Secondly, we consider in Section~\ref{sec3} how the structure of the underlying social network and the modelling of user-to-user transmission mechanisms can affect the offspring distribution for cascade trees. By comparing with the empirical results of Sec.~\ref{sec2}, we examine whether the offspring distribution is better modelled by the independent cascade model, or by an alternative model that accounts for limited attention of users of social media \cite{Lerman16}.

Finally, in Section~\ref{sec4}, we use the branching process framework to derive predictions for features of cascades, focusing on both the distribution of metrics of interest across the entire dataset and on analytical results for expected values. For completeness, we first derive the well-known results for cascade sizes and durations and then build on this approach to derive results for the expected value of tree depths and the structural virality measure of Goel et al.~\cite{Goel15}, and apparent novelty decay factors  \cite{WuH07,Yasseri17}. These results include integral expressions for expected structural virality and tree depth that we have not been able to locate in existing literature, and which are amenable to asymptotic analysis.
We conclude the paper with a discussion of the results, limitations, and potential extensions in Section~\ref{sec5}.


\section{Data} \label{sec2}

\subsection{Data sources} \label{sec2.1}
As a motivation and test for branching process hypotheses, we reanalyse two independent Twitter datasets. Both datasets have been previously analysed and described in Refs.~\cite{OSullivan17} and \cite{Lerman12,Hodas14}, but here we use the identified cascade structures to focus on the accuracy of branching process models.

The first dataset, which we call ``Marref'', is comprised of tweets related to the 2015 Irish same-sex marriage referendum, collected between May 8 and May 23, 2015. As described in \cite{OSullivan17}, all tweets containing either of the hashtags \#marref or \#marriageref were gathered and analysed. Our focus in this paper is on the tree structures of retweeting behaviour, and the extent to which this can be accurately modelled by branching processes. A ``particle'' or ``node'' in a tree (see Figure~\ref{figT1} for examples) represents a retweet event, which may cause further retweets, i.e., a next generation of particles in the tree, called the ``children'' particles of the ``parent'' node. The children of a given node event are identified using the tree-reconstruction methodology described in Goel et al.~\cite{Goel15}, as implemented in \cite{OSullivanPhDthesis}. This algorithm aims to identify a single parent node (tweet) for each retweet event, by using the text of the tweet and (in cases of multiple possible parents) the timestamps of the tweeting events. The output of the Goel et al.~algorithm is a tree structure corresponding to each cascade of retweets, as described in Sec.~\ref{sec2.2} below. The data collection procedure is restricted to cascades of size greater than one, so that seeds that generate no children are not recorded.

The second dataset, called ``URL'', is comprised of tweets containing Uniform Resource Locators (URLs) for internet addresses that were posted on Twitter during October 2010 \cite{LermanDataWebsite}. The collection and processing of the data is described in detail in \cite{Lerman12} and \cite{Hodas14}. The URLs chosen for tracking were discovered by sampling tweets gathered using Twitter's Gardenhose API \cite{Hodas14} and then searching for all appearances of these URLs. As for the Marref dataset, trees were reconstructed using the algorithm of Goel et al.~\cite{Goel15}. 
We note that there is some selection bias is this data as URLs with larger cascade sizes were more likely to be selected by the initial sampling than relatively unpopular URLs. Also, as in the Marref data, the URLs dataset only contains those cascades whose size (total number of tweets) is greater than one.

In order to ensure that each of the studied trees has had sufficient time to develop all its generations within the finite-length time window of data collection, we select only those trees where the ``birth'' of the tree (i.e., the timestamp of the seed node) occurs during the first half of the time window for data collection. In the case of the Marref data, this avoids trees that are born during the increased activity around the referendum day and its aftermath.
Although the original tweets occur in continuous time, we confine our attention in this paper to the discrete generations of the tree structures. This choice simplifies the analysis to the discrete-time case of Galton-Watson processes, but it means that we do not investigate important aspects such as the time between tweet arrival and retweet, which would require an extension of our approach to continuous-time branching processes \cite{Iribarren09,Iribarren11}.
The extracted tree structures for both datasets are made available in anonymised form \cite{GithubTrees}.

\subsection{Tree structures} \label{sec2.2}
\begin{figure}
\centering
\epsfig{figure=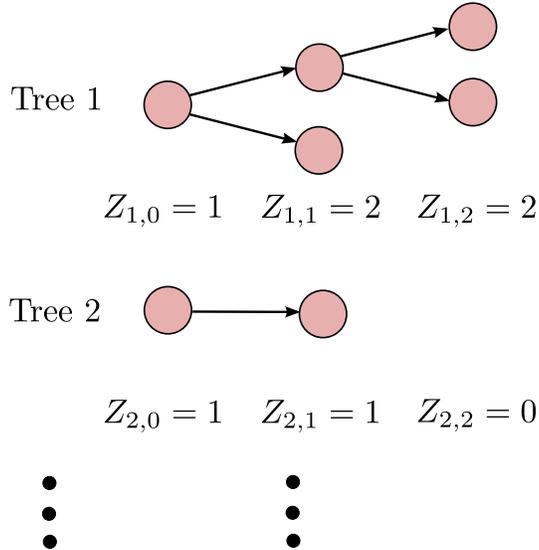,width=7cm}
\caption{Schematic of the ensemble of trees, indicating the $Z_{m,n}$ values for the first two trees in the ensemble. }\label{figT1}
\end{figure}

In this section we analyse the characteristics of the trees extracted from the data as described in Sec.~\ref{sec2.1}. For each dataset, we consider an ensemble of $M$ trees, with each tree made up of particles (or nodes) in multiple generations, see Fig.~\ref{figT1}.
We define $Z_{m,n}$ to be the number of particles in generation $n$ of tree $m$ (where $m=1,2,\ldots,M$ and  $n=0,1,\ldots,$ ). The individual trees have very heterogeneous characteristics (size, number of generations, etc.), so we first consider the ensemble as a whole.

\begin{figure}
\centering
\epsfig{figure=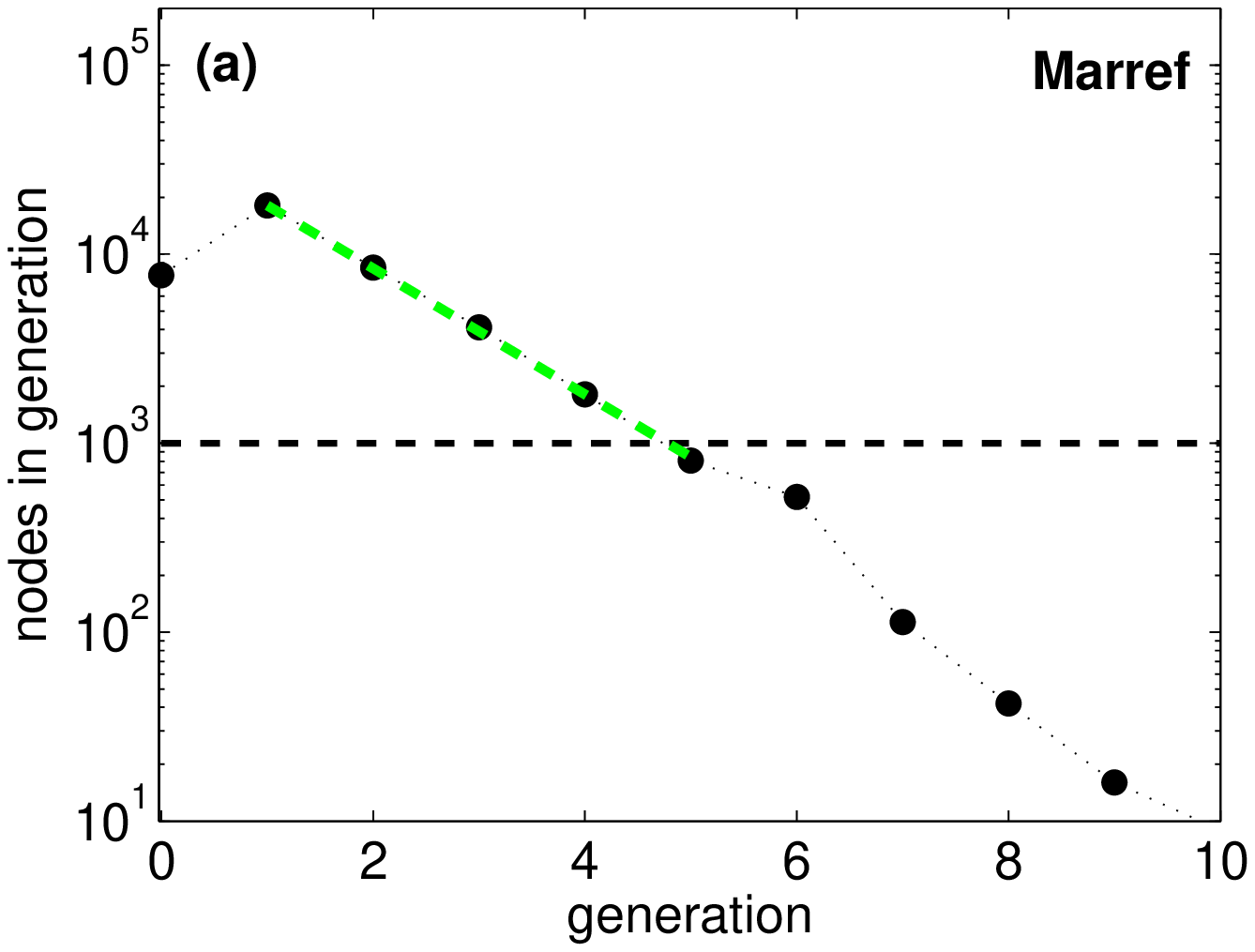,width=8.1 cm} 
\epsfig{figure=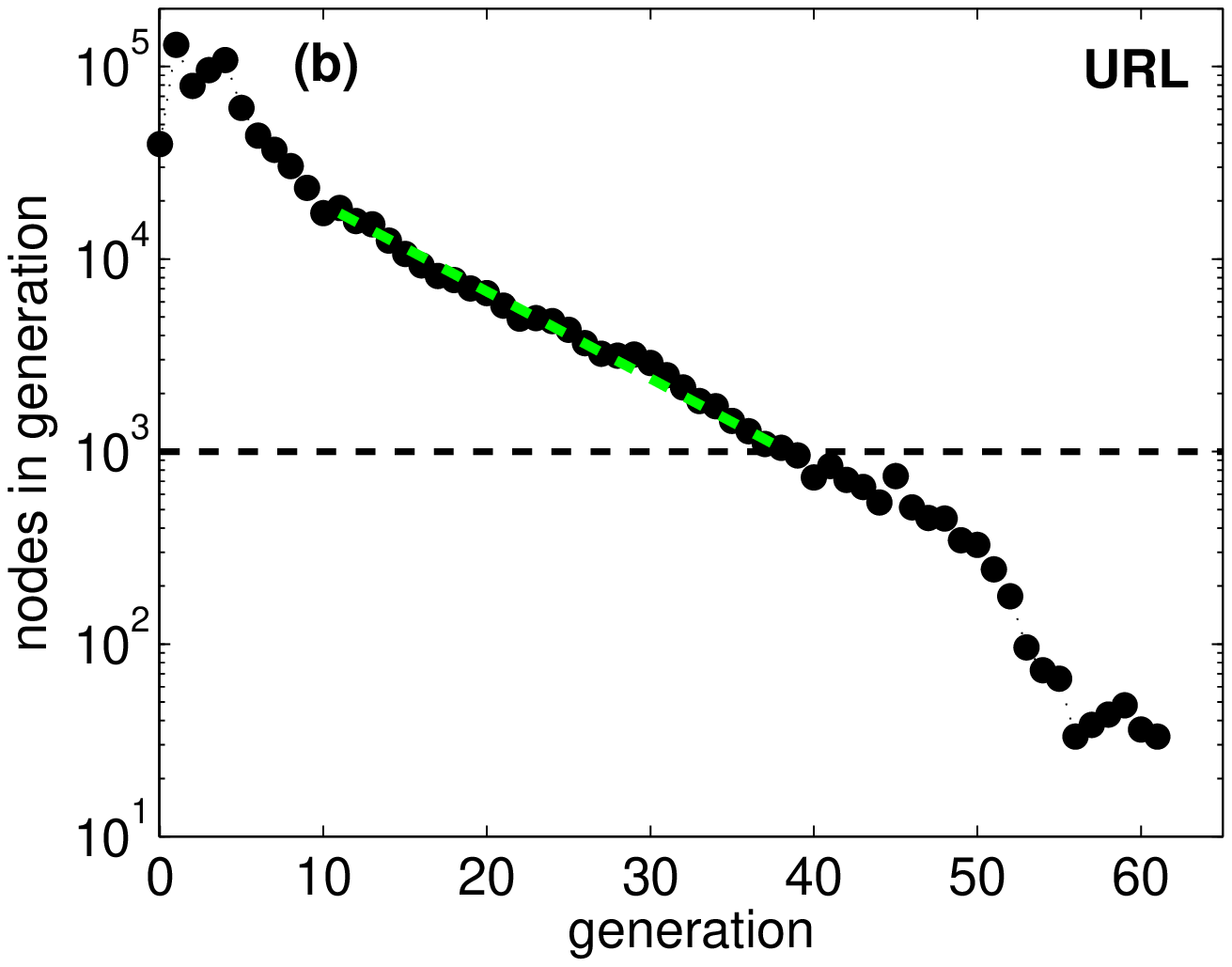,width=8.1 cm}\\
\epsfig{figure=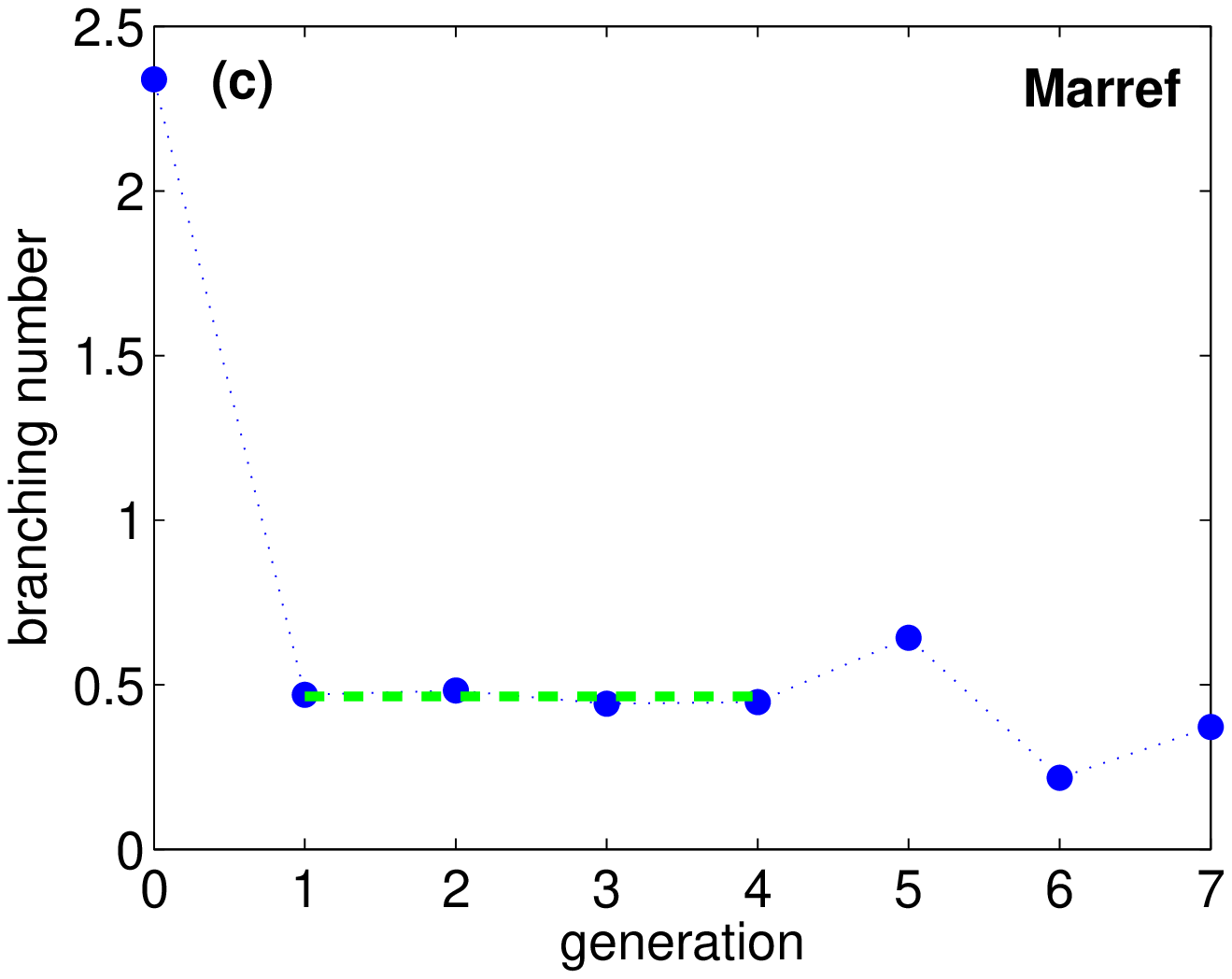,width=8.1 cm}
\epsfig{figure=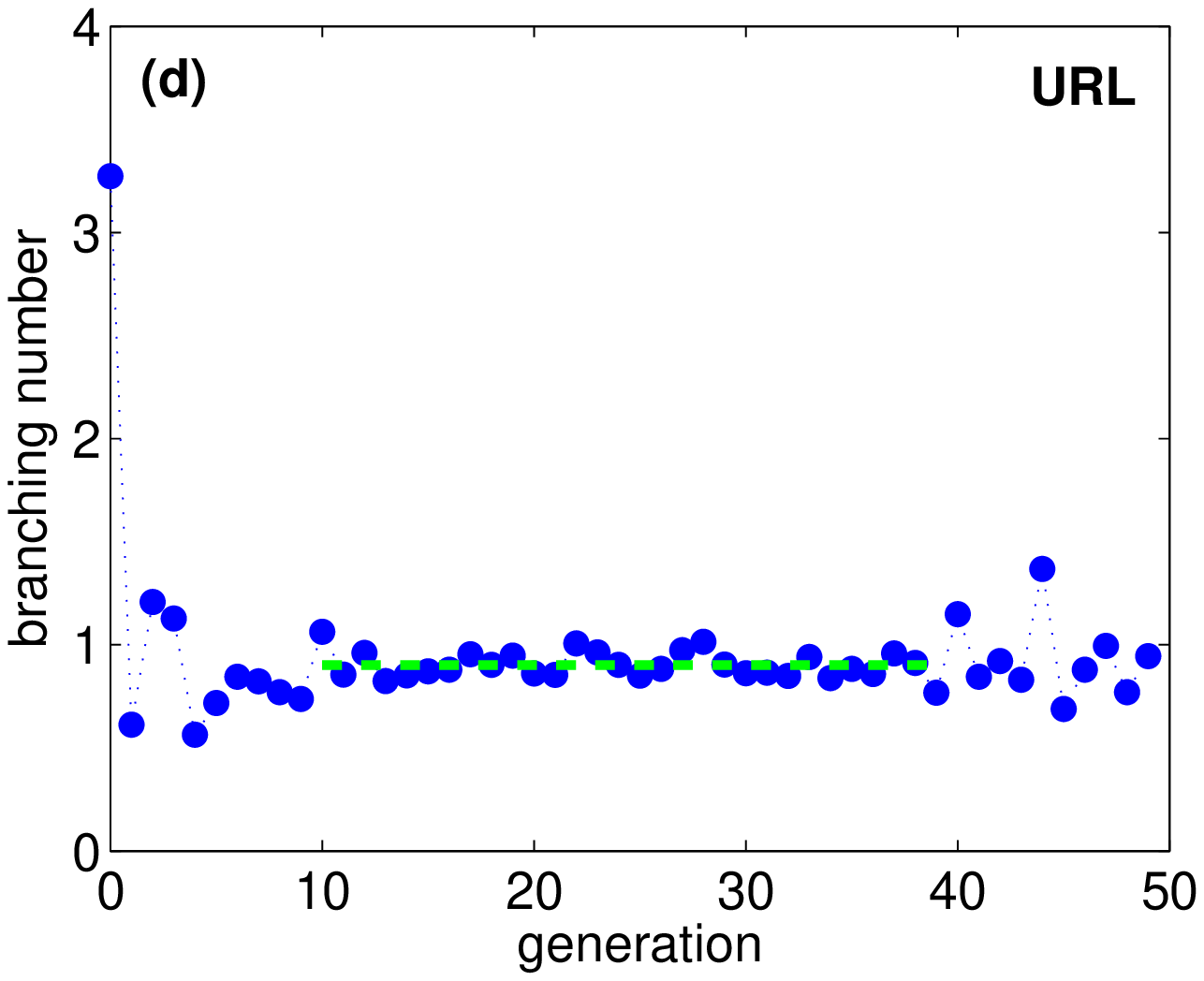,width=8.1 cm}
\caption{Number of nodes (top panels) and effective branching number (bottom panels) in the data, as defined by Eqs.~(\ref{zneq}) and (\ref{xineq}). Here, and in most subsequent figures, the left panels ((a) and (c)) show results for the Marref dataset while the right panels ((b) and (d)) are the results from the URL dataset.}\label{fig1}
\end{figure}

Defining $z_n$ as the total number of generation-$n$ particles observed across all trees, i.e.,
\begin{equation}
z_n = \sum_{m=1}^M Z_{m,n}, \label{zneq}
\end{equation}
we plot in the top panels of Fig.~\ref{fig1} the dependence of $z_n$ on the generation $n$ using log-linear scales. Figure~\ref{fig1}(a) is from the Marref dataset ($M=7,736$) and Fig.~\ref{fig1}(b) is from the URL dataset ($M=39,547$).  An approximately exponential dependence of $z_n$ upon $n$ is shown by the nearly linear shape of the function on log-linear axes; such a dependence is consistent with a subcritical branching process.
Note that small-number fluctuations occur when $z_n$ is relatively small: we choose $10^3$ as a threshold level (shown by the black dashed line in Figs.~\ref{fig1}(a) and (b)) and focus on $z_n$ values which are above this threshold.

In Fig.~\ref{fig1}(c) and Fig.~\ref{fig1}(d) (for Marref and URL, respectively) we show the \emph{effective branching number} \cite{Dobson12}
\begin{equation}
\xi_n = \frac{z_{n+1}}{z_n}, \label{xineq}
\end{equation}
which gives the average number of children particles for a particle of the $n$th generation. Observe that $\xi_n$ is approximately constant for the range of generations in which $z_n$ is sufficiently large (i.e., above the threshold marked in Figs.~\ref{fig1}(a) and (b)). Figure~\ref{fig1}(d) shows that  the early generations of the URL dataset exhibit a lot of fluctuations in the $\xi_n$ values, consistent with the possible biasing of the data towards larger trees (see Sec.~\ref{sec2.1}). In both cases, the branching number $\xi_0$ of the seed generation appears to be anomalously high;  this is partly due to the biasing introduced by the fact that no trees of size less than two are recorded (but see also the discussion leading to Eq.~(\ref{3.7}) below). The dashed green lines in Fig.~\ref{fig1} highlight the range of generations over which the branching number $\xi_n$ appears to be approximately constant. For each dataset we calculate an average value $\bar\xi$ of the branching number over the range shown by the dashed green line. The URL dataset, with a value $\bar \xi=0.90$, has a high virality (recall the critical branching number of 1 separates the regime of subcritical cascades from that of supercritical cascades), while the Marref dataset has lower virality ($\bar \xi = 0.46$).

Next, we make a stronger test of the branching process hypothesis, by examining the empirical offspring distribution at each generation. For each particle $i$ in generation $n$ we record the number $Z_{n+1}^{(i)}$ of its offspring particles, i.e., the number of users in generation $n+1$ that are identified as children of particle $i$. Gathering the ensemble of $Z_{n+1}^{(i)}$ values across all trees, we calculate the empirical offspring distribution of generation $n$ as
\begin{equation}
\bar{q}_{\ell,n}= \text{Prob}\left(  Z_{n+1}^{(i)}= \ell \left| \text{ particle }i\text{ in generation }n\right. \right),
\end{equation}
i.e., $\bar{q}_{\ell,n}$ is the probability that a particle in generation $n$ spawns $\ell$ children particles in generation $n+1$ and we have used the fact that the maximum-likelihood estimate of the probability of having $\ell$ children is given by the fraction of nodes in the data with $\ell$ children \cite{Golub10}.

\begin{figure}
\centering
\epsfig{figure=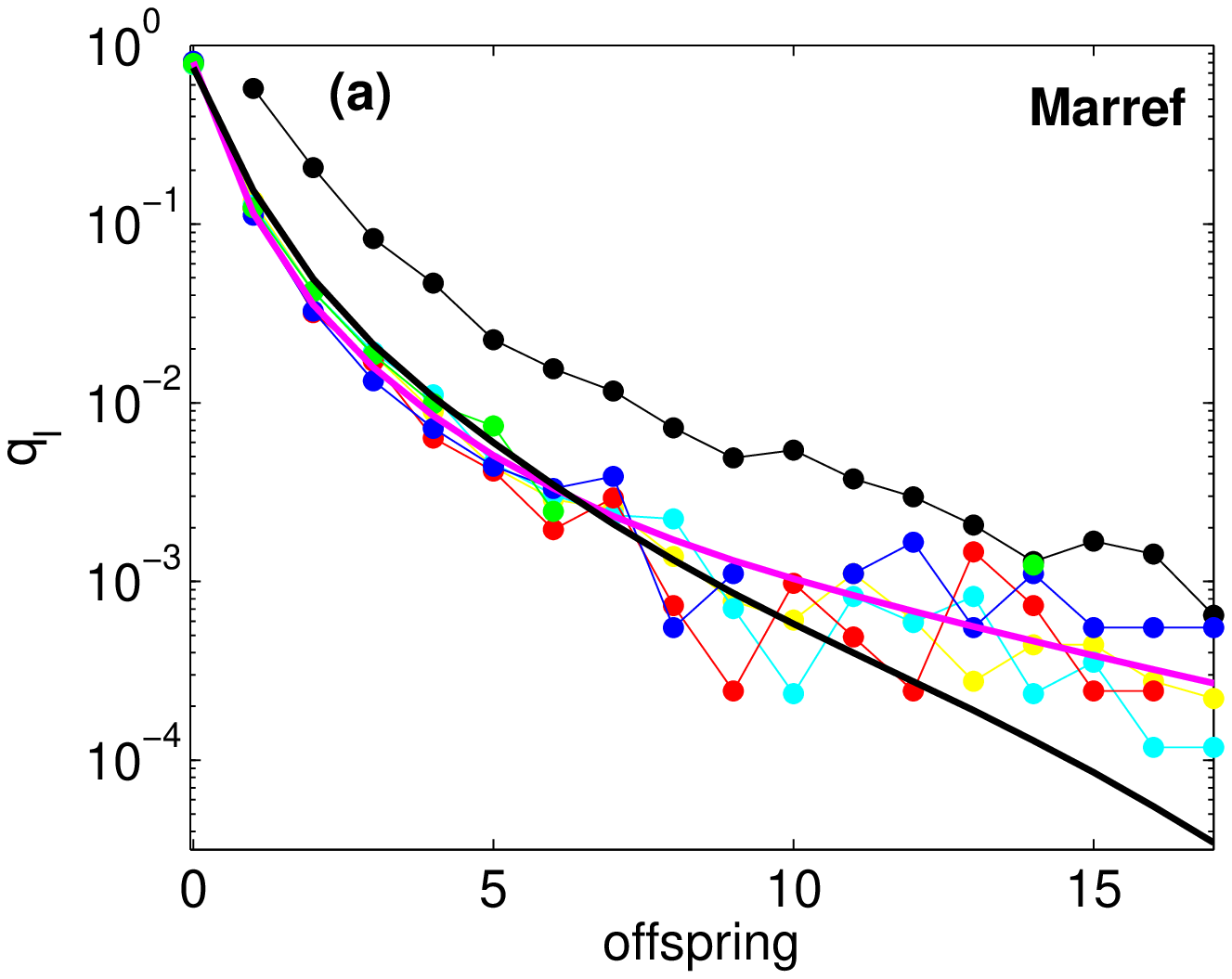,width=8.1 cm}
\epsfig{figure=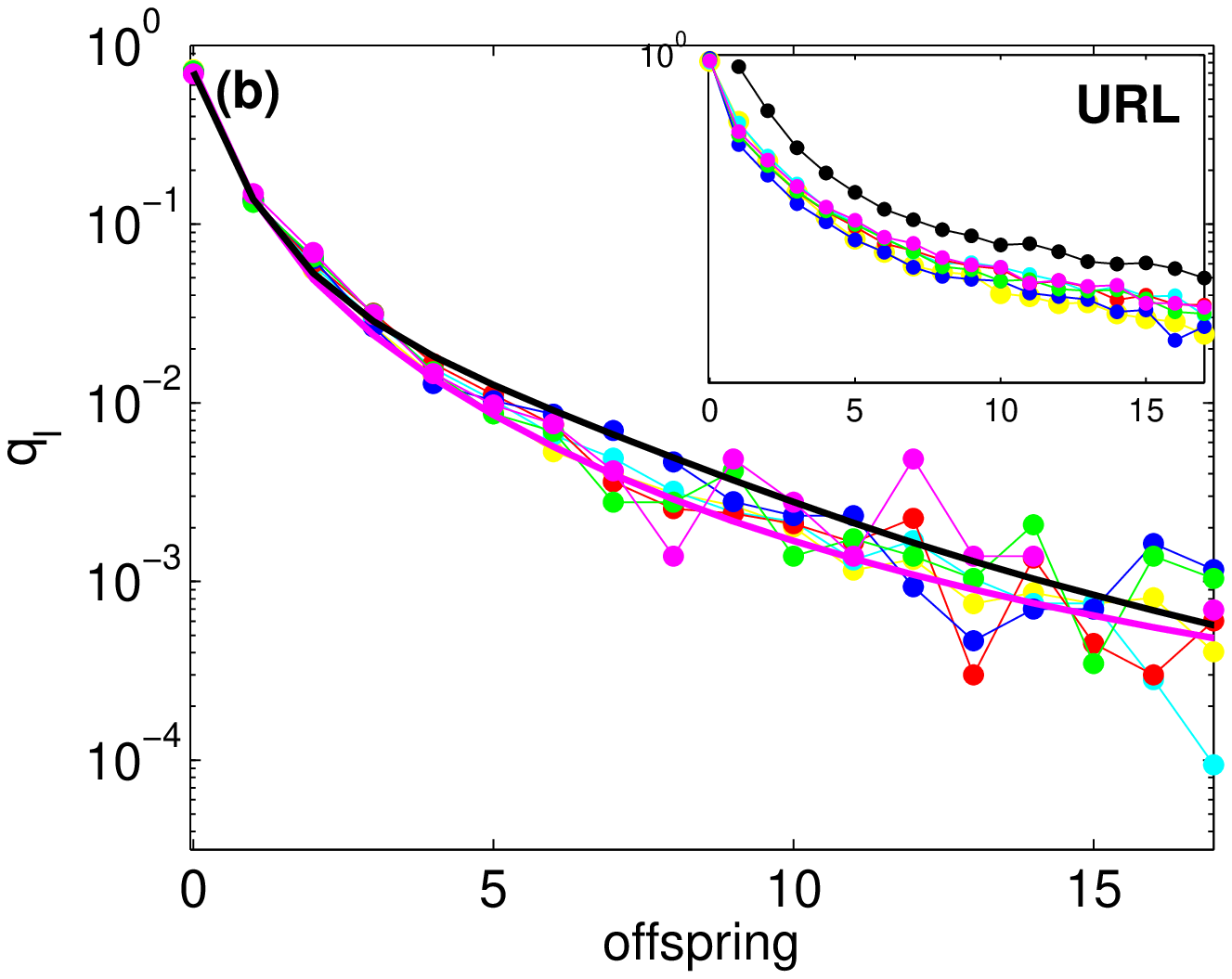,width=8.1 cm} \\
\epsfig{figure=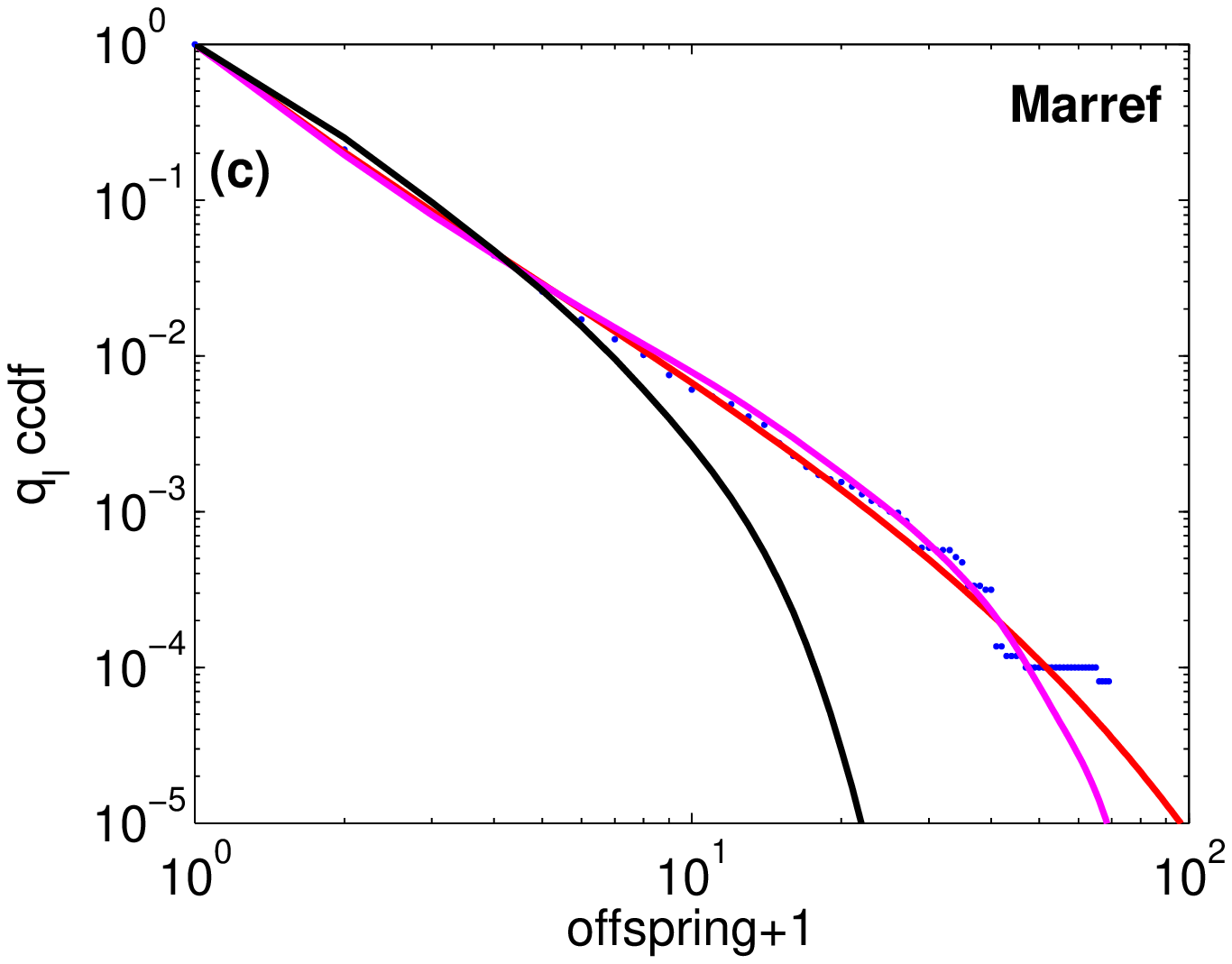,width=8.1 cm}
\epsfig{figure=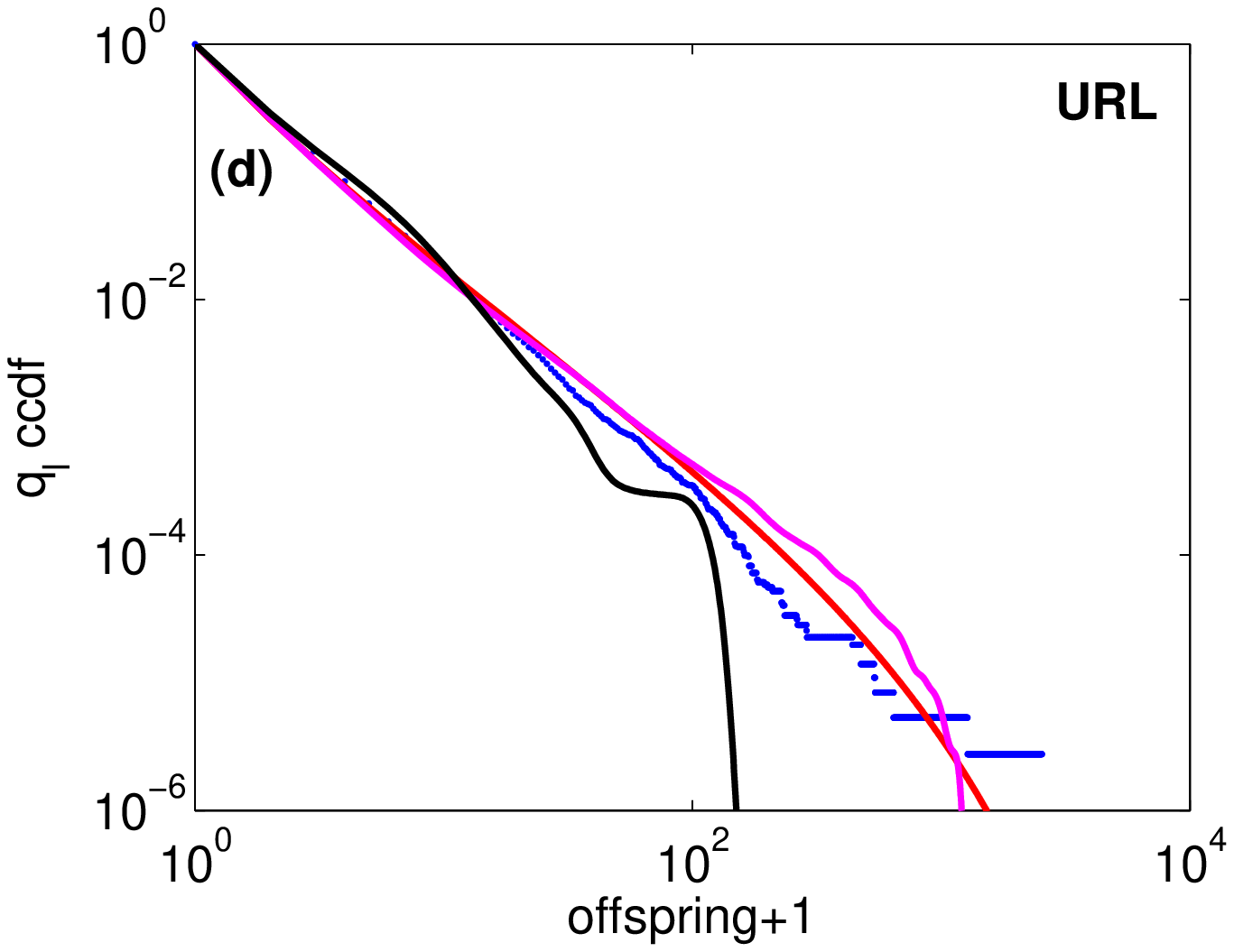,width=8.1 cm}
\caption{Empirical offspring distributions. (a) Offspring distributions for generations 0  (black symbols) through to 4 (coloured symbols) of Marref dataset. The magenta curve is the LAM prediction; the black curve is the ICM prediction. (b) Offspring distributions for generations 10 through to 40 in steps of 5 (with generations 0 to 5 in inset) from URL dataset, with LAM and ICM theory curves in magenta and black, respectively. (c) CCDF of offspring distribution for Marref; blue symbols show the averaged empirical distribution (averaged over generations 1 through 4), curves are LAM (magenta) and ICM (black) predictions, with the fitted distribution of Eq.~(\ref{3.15}) in red. (d) As panel (c), but for the URL datset, with the averaged empirical distribution averaged over generations 10 through 40. 
}\label{fig2}
\end{figure}

In Figs.~\ref{fig2}(a) and (b) we plot the empirical offspring distributions for several generations. Because of the data collection restriction to cascades of size exceeding one (and also because of the network structure, see Sec.~\ref{sec3} below), the seed generation offspring distribution $\bar{q}_{\ell,0}$ differs substantially from the other generations. However, for the Marref data set (Fig.~\ref{fig2}(a)), observe that the $\bar{q}_{\ell,n}$ distributions for $n=1$ through $n=4$ (which is the range of generations giving $z_n$ values above threshold in Fig.~\ref{fig1}(a)) are very similar to each other: the curves in Fig.~\ref{fig2}(a) are almost indistinguishable. This collapse of the empirical offspring distributions is consistent with a branching process model in which the offspring distributions are identical for all generations with $n\ge 1$, see Sec.~\ref{sec3}.

In the URL data set, the low-generation distributions $\bar{q}_{\ell,n}$ do not show as  clean a collapse as seen in the Marref case, see the inset of Fig.~\ref{fig2}(b). However, this may be due to the selection bias  in the data collection, which means that small trees (those with fewer generations) are likely to have been omitted from the collected set of trees. Larger trees are more likely to be properly represented in the dataset, and these trees are also likely to consist of a large number of generations. Accordingly, we plot also the $\bar{q}_{\ell,n}$ curves for $n=10, 15, 20, 25, 30, 35, 40$ in Fig.~\ref{fig2}(b) (note the range of generations chosen matches the green dashed line in Fig.~\ref{fig1}(b) and (d)), and we observe a good collapse of these distributions, which is again consistent with a branching process model.

From the evidence of Figs.~\ref{fig1}, \ref{fig2}(a) and \ref{fig2}(b), we conclude that a branching process model may give a good approximation to the heterogeneous cascades represented by the trees extracted from the data sets. In the next section we will derive a mathematical model that explicitly links the network structure and  various hypotheses on the information-spreading mechanism to predict offspring distributions, which we then compare with the empirical results of Fig.~\ref{fig2}(a) and \ref{fig2}(b).

\section{Modelling information spread as a branching process} \label{sec3}

We consider a directed network whose structure is minimally described (in the configuration-model sense \cite{Newmanbook}) by the joint distribution $p_{j k}$ of nodes' in-degree $j$ and out-degree $k$: in other words, $p_{j k}$ is the probability that a randomly chosen node has $j$ friends and $k$ followers\footnote{Following \cite{Lerman12,Hodas14}, we call the nodes followed by node $i$ (the in-neighbours of $i$) its \emph{friends}. The \emph{followers} of $i$ are the out-neighbours of $i$, where we consider the direction of the edges to be the direction of information flow, i.e., edges point from a node to its followers.}. We model the dynamics of information spreading at the level of $(j,k)$ classes also, defining the \emph{vulnerability} $v_{j k}$ as the probability that a $(j,k)$-class node will retweet a message that it has received from one of its $j$ friends \cite{GleesonDurrett}.  

Consider a message that is tweeted by a node to its followers. Under the configuration-model assumption, 
 the probability that a follower is in the $(j,k)$ class is given by $j/\left< j\right> p_{j k}$, where $\left< j\right> = \sum_{j,k} j p_{j k}$ is the mean in-degree (mean number of friends) over the network. This follower will retweet the message if he is vulnerable, which occurs with probability $v_{j k}$, and in doing so, he will expose all $k$ of his followers to the message. Thus, the probability that a randomly-chosen follower will retweet a message he receives is given by
\begin{equation}
\rho = \sum_{j,k} \frac{j}{\left< j\right>} p_{j k}v_{j k}. \label{3.1}
\end{equation}
If we know that a follower has retweeted the message (i.e., if we condition on retweeting) then the probability that he is in the $(j,k)$ class is
\begin{equation}
\frac{1}{\rho} \frac{j}{\left< j \right>} p_{ j k} v_{j k}. \label{3.2}
\end{equation}
In particular, the probability that a retweeter has $k$ followers is given by summing over all possible $j$ values:
\begin{equation}
\sum_j \frac{1}{\rho} \frac{j}{\left< j \right>} p_{ j k} v_{j k}. \label{3.3}
\end{equation}
Assuming each of the $k$ followers to be independently vulnerable with probability $\rho$, the number $\ell$ of  followers who themselves retweet has the binomial distribution
\begin{equation}
{{k}\choose{\ell}} \rho^\ell (1-\rho)^{k-\ell}. \label{3.4}
\end{equation}
Combining these probabilities, we have derived the offspring distribution $q_\ell$ which gives the probability that a retweeting by a node will lead to $\ell$ further retweets by followers of that node as
\begin{equation}
q_\ell = \sum_k \underbrace{\sum_j \frac{1}{\rho} \frac{j}{\left< j\right>} p_{j k}v_{j k}}_{\text{Prob }k\text{ followers, conditioned on retweeting}}\underbrace{{{k}\choose{\ell}} \rho^\ell (1-\rho)^{k-\ell}}_{\text{Prob }\ell\text{ of }k\text{ followers retweet}} .\label{3.5}
\end{equation}
The corresponding pgf for the offspring distribution, $f(x) = \sum_\ell q_\ell x^\ell$, is
\begin{equation}
f(x) = \sum_{k,j}\frac{1}{\rho}\frac{j}{\left< j \right>} p_{j k} v_{j k}\left( 1- \rho+ \rho x\right)^k. \label{3.6}
\end{equation}

In the derivation of Eq.~(\ref{3.6}), we began by considering a node that receives the message from one of its friends. However, the initial source (or \emph{seed}) of the cascade has a different dynamic, meaning that the seed generation of the branching process has an offspring distribution different from Eq.~(\ref{3.5}). We assume that the seed node for a cascade is chosen uniformly at random from all the nodes. This means that the seed node is in the $(j,k)$ class with probability $p_{j k}$. As above, the number $\ell$ of its $k$ followers who will retweet the message is given by Eq.~(\ref{3.4}), and so the seed-generation offspring distribution is\footnote{We use tildes to differentiate the seed-generation offspring distribution and pgf from those defined in Eqs.~(\ref{3.5}) and (\ref{3.6}).}
\begin{equation}
\widetilde{q}_{\ell} =\sum_{k,j} p_{j k} {{k}\choose{\ell}} \rho^\ell(1-\rho)^{k-\ell}, \label{3.7}
\end{equation}
with corresponding pgf
\begin{equation}
\widetilde{f}(x)=\sum_{\ell=0}^\infty \widetilde{q}_{\ell} x^\ell = \sum_{k, j} p_{j k}\left( 1-\rho+\rho x\right)^k. \label{3.8}
\end{equation}

Having defined how the offspring distribution of the branching process is determined by the network structure ($p_{ j k}$) and the dynamics (via the vulnerability $v_{j k}$), we next examine two possible models for contagion dynamics.

\subsection{The independent cascade model} \label{ICM}
In the independent cascade model (ICM) \cite{Kempe03} each ``infected'' node (i.e., node who tweets or retweets the message of interest) gets one attempt to infect each of its out-neighbours; the infection attempt is successful (meaning that the follower also retweets the message) with probability $C$, where $C$ is the single parameter of the model. In our modelling framework, this implies that the ICM vulnerability of every node is equal to $C$, regardless of the node's $(j,k)$ class:
\begin{equation}
v_{j k}^\text{ICM} = C\quad \forall j,k. \label{3.9}
\end{equation}
Note that in this case, the retweet probability $\rho$ is determined from Eq.~(\ref{3.1}) to be $\rho=C$. Moreover, in the special case of uncorrelated in- and out-degrees (i.e., if the number of friends $j$ and the number of followers $k$ of a node are uncorrelated), the joint distribution $p_{j k}$ factorises into the product $p_j^\text{in} p_k^\text{out}$ and the offspring $q_{\ell}$ and $\widetilde{q}_\ell$ are identical\footnote{This is easily seen from the corresponding pgfs, where the sum over $j$ can be performed in Eqs.~(\ref{3.6}) and (\ref{3.8}) to give $f^\text{ICM}(x)=\widetilde{f}^\text{ICM}(x) = \sum_k p_{k}^\text{out}\left(1-C+C x\right)^k$ if $p_{j k} = p_j^\text{in} p_k^\text{out}$.}. However, in the more realistic case where the in- and out-degrees of nodes (the numbers of friends and followers of users)  are correlated (see, for example, Fig.~2 of \cite{Ma18}), the offspring distribution of the seed generation differs from the offspring distributions of subsequent generations.

\subsection{The limited attention model} \label{LAM}

A number of researchers have pointed out that the limitations of human cognition impose an effective limit on how much information can be absorbed and shared by an individual. For a user on Twitter, having a larger number of friends $j$ leads to a faster influx of information into the user's stream, with a consequent dividing of attention among the many tweets. Empirical analyses \cite{Hodas14,Lerman16} and models of information-sharing dynamics \cite{Weng12,GleesonPRL14,Gleeson16} both indicate that the probability that a user retweets a particular piece of information she has received can be modelled as being approximately inversely proportional to the number $j$ of her friends. In our notation, the vulnerability $v_{j k}$ of a $(j,k)$-class user in the limited attention model (LAM) is inversely proportional to $j$:
\begin{equation}
v_{j k}^\text{LAM} = \frac{B}{j}, \label{3.11}
\end{equation}
where $B$ is a parameter of the model, and we assume no nodes have $j=0$.

In the LAM, the probability $\rho$ of a random follower retweeting is given in terms of $B$ by Eq.~(\ref{3.1}):
\begin{equation}
\rho=\frac{B}{\left< j \right>}. \label{3.12}
\end{equation}
Interestingly, under the assumption that the network has no nodes with $j=0$ then the LAM offspring distributions for the seed generation and for later generation are identical, even if the in- and out-degrees of nodes are correlated (unlike the ICM model):
\begin{equation}
f^\text{LAM}(x) = \widetilde{f}^\text{LAM}(x) = \sum_{k,j} p_{j k} \left( 1- \rho + \rho x\right)^k \quad \text{ if }p_0^\text{in}=0. \label{3.13}
\end{equation}

\subsection{Comparing ICM and LAM with empirical offspring distributions}
Using the empirical network structure for the Marref and URL datasets, specifically the in-degree $j_i$ and out-degree $k_i$ of each node $i$ in the network, we construct the offspring distribution predicted by the independent cascade model and by the limited attention model, using Eqs.~(\ref{3.9}) and (\ref{3.11}), respectively, in Eqs.~(\ref{3.1}), (\ref{3.5}) and (\ref{3.7}). In each case, we fit the parameters $C$ and $B$ by matching the branching number to the average value $\bar \xi$ calculated in Sec.~\ref{sec2.2}. The sums over $j$ and $k$ are replaced by sums over the $N$ nodes: Equation~(\ref{3.1}), for example, becomes
\begin{equation}
\rho= \sum_{i=1}^N \frac{j_i}{\bar{j}}\frac{1}{N} v_{j_i k_i},
\end{equation}
where $\bar j$ is the sample mean of the in-degrees: $\bar j = \frac{1}{N}\sum_{i=1}^N j_i$. (In effect, we replace $p_{j k}$ by $1/N$ and replace sums over $j$ and $k$ by a sum over all nodes.)

The black (for ICM) and magenta (for LAM) curves in Fig.~\ref{fig2} show how these predictions compare with the empirical offspring distribution.
Evidently, the LAM predictions are closer to the empirical offspring distributions than the ICM predictions, at least for the relatively low values of $\ell$ in Figs.~\ref{fig2}(a) and (b). To examine the empirical offspring distributions at higher values of $\ell$ we reduce the low-number fluctuations by averaging the distributions over the generations marked with the green line in Fig.~\ref{fig1}(a) and (b), i.e., those generations for which the effective branching number is approximately constant. This averaged offspring distribution is shown by the blue symbols in Figs.~\ref{fig2}(c) and (d): note we plot $\ell+1$ on the horizontal axis  in order to make the $\ell=0$ case visible on the logarithmic scale.

Noting the near-linear decay of the offspring distribution on the log-log plot, we fit the empirical averaged offspring distribution with a truncated power law:
\begin{equation}
q_\ell \propto \left(\ell+1\right)^{-\beta} e^{-\frac{\ell}{\theta}}.\label{3.15}
\end{equation}
This distribution is chosen for its good fit and analytical convenience\footnote{Its pgf is $\text{Li}_{\beta}\left(e^{-\frac{1}{\theta}} x\right)/\left(x \text{Li}_{\beta}\left(e^{-\frac{1}{\theta}}\right)\right)$, where $\text{Li}_\beta$ is the polylogarithm function of order $\beta$.}; calculations with this distribution can be more easily reproduced than by using the full ICM or LAM distributions, which require knowledge of the full set of node degrees $(j_i,k_i)$. To fit the parameters $\beta$ and $\theta$ in Eq.~(\ref{3.15}), we match the first and second moments of the distribution with the corresponding moments of the averaged empirical distribution. The fitted parameters are given in Table~\ref{tab1}, and the red curves in Figs.~\ref{fig2}(c) and \ref{fig2}(d) show that the fitted offspring distribution is reasonably close to the empirical distribution. A similar procedue is used to fit a seed generation offspring distribution $\widetilde{q}_\ell$, using the form of Eq.~(\ref{3.15})  with parameters $\beta$ and $\theta$ replaced by $\beta_0$ and $\theta_0$, and with the domain restricted to $\ell>1$ because every seed node in the empirical dataset has at least one child\footnote{The pgf for the seed generation is $\left(x-e^{\frac{1}{\theta}} \text{Li}_{\beta}\left(e^{-\frac{1}{\theta}} x\right)\right)/\left(x-x e^{\frac{1}{\theta}}   \text{Li}_{\beta}\left(e^{-\frac{1}{\theta}}\right)\right)$.}.
\begin{table}
\begin{center}
\begin{tabular}{|c||c|c|}
\hline
    & Marref & URL  \\
\hline\hline
$\beta$ & 2.72 & 2.48  \\ \hline
$\theta$ & 47.6& 1055 \\ \hline
$\beta_0$ & 2.82 & 2.58 \\ \hline
$\theta_0$ & 178 & $1.82 \times 10^5$\\
 \hline
\end{tabular}
\end{center}
\caption{Parameter values for the distribution in Eq.~(\ref{3.15}), fitted to the first and second moment of the averaged empirical distributions.}
\label{tab1}
\end{table}

To summarize this Section: we have derived a general formulation for the offspring distribution that results from cascades on a network with a given distribution $p_{j k}$ of in- and out-degrees. We used the vulnerability $v_{j k}$ to describe different models of information transmission, focussing on comparing the ICM with the LAM. In Fig.~\ref{fig2} we see that there are observable differences between the offspring distributions predicted by the two models, with the LAM case generally closer to the empirical observations. Finally, we fitted a standard distribution (Eq.~(\ref{3.15})) to the empirical distribution to make our results in the next section more tractable and readily reproducible. Note, however, that in principle the data on the structure of the network (e.g., the $p_{j k}$ distribution) and the assumed vulnerability $v_{j k}$ suffice to determine the offspring distribution, and this opens the possibility of examining further hypotheses on the dependence of information spreading on the nodes' in- and out-degrees \cite{Wu18}.

It is also worth noting that the network structure, through the correlations in the $p_{j k}$ distribution, strongly affects the offspring distribution (and hence, as we show in the next Section, the predictions of the cascade structure); this point has recently been recognised by Ma et al.~\cite{Ma18}. We point out that the $p_{j k}$ distribution of the network should therefore be included, when possible, in analysis of information spreading. This is not current practice: in Refs.~\cite{Goel15,Gleeson16}, for example, large-scale simulations  are performed on synthetic networks with specified out-degree distributions but without considering the correlation structure between the in- and out-degrees of nodes.

\section{Application of branching process model}\label{sec4}
In this Section we focus on analytical predictions of branching process theory that can be compared to statistical features of the two datasets. We begin with a discrete-time branching process, where---as in Sec.~\ref{sec3}---the number of offspring of the seed particle is distributed according to pgf $\widetilde{f}(x) = \sum_{\ell=0}^\infty \widetilde{q}_\ell x^\ell $ while all later generations of the tree have offspring numbers generated by $f(x) = \sum_{\ell=0}^\infty q_\ell x^\ell$. We consider the seed of the tree to be generation 0, and we are interested in various properties of the trees as observed a number of generations later. In Sections \ref{sec4.1} and \ref{treesize} we use a slightly unusual approach to derive known results on the distribution of cascade durations and sizes. We then extend this methodology to the calculation of other metrics in Secs.~\ref{sec4.3} and \ref{novelty}.

\subsection{Number of particles in generation $n$; distribution of cascade lifetimes}\label{sec4.1}
\begin{figure}
\centering
\epsfig{figure=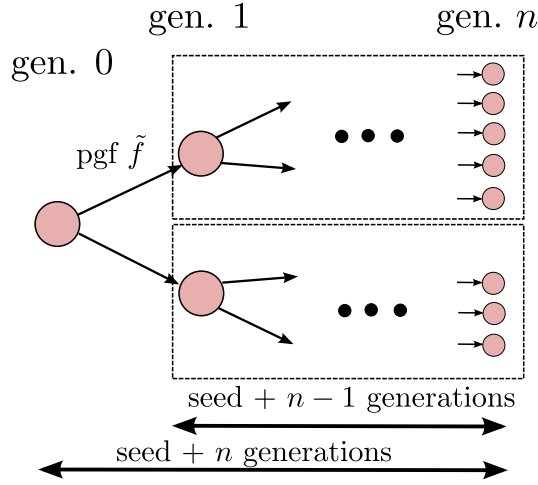,width=7cm}
\caption{Schematic of a tree generated by a seed particle; note the number of children of the seed particle is generated by $\widetilde{f}$.}\label{schematic_subtree1}
\end{figure}

As a first example, we define the random (non-negative integer) variable $\widetilde{Z}_n$ to represent the number of particles in generation $n$ of the tree (the small nodes in Fig.~\ref{schematic_subtree1}). As schematically represented in Fig.~\ref{schematic_subtree1}, these particles are the descendants of the generation-$0$ seed node, observed $n$ generations after the seed. They   can also be considered as the sum of the particles contained in all the subtrees that are seeded at generation 1 and which are observed $n-1$ generations later. Conditioning on the number $k$ of particles in generation 1, we define $Z_{n-1}^{(i)}$ to be the number of particles in the subtree that is seeded by the $i$th particle in generation 1, as observed $n-1$ generations after the subtree is born (i.e., at generation $n$ of the parent tree).  Since all the subtrees are i.i.d., each of the $k$ random variables $Z_{n-1}^{(i)}$ has the same distribution.

We define the pgf $\widetilde{F}_n$ for the random variable $\widetilde{Z}_n$ as
\begin{equation}
\widetilde{F}_n(s) = E\left(s^{\widetilde{Z}_n}\right) = \sum_{j=0}^\infty \text{Prob}\left(\widetilde{Z}_n =  j\right) s^j \label{1},
\end{equation}
where $E$ denotes expectation over the ensemble of trees and $s$ is a dummy variable. If we condition on the number $k$ of particles in generation 1, we can write $\widetilde{Z}_n$ as the sum of the $k$ subtree variables $Z_{n-1}^{(i)}$ (the superscript $i$ denotes the $i$th i.i.d. copy):
\begin{equation}
\widetilde{Z}_n = \sum_{i=1}^k Z_{n-1}^{(i)}, \label{2}
\end{equation}
and so
\begin{align}
E\left( \left. s^{\widetilde{Z}_n} \right|\, k\text{ particles in generation 1}\right) & = E\left( s^{\sum_{i=1}^k {Z_{n-1}}^{(i)}}\right)\nonumber \\
&= E\left(s^{Z_{n-1}^{(1)}}\right) E\left(s^{Z_{n-1}^{(2)}}\right)\ldots E\left(s^{Z_{n-1}^{(k)}}\right)\nonumber \\
&= \left[ E\left(s^{Z_{n-1}}\right)\right]^k,\label{3}
\end{align}
where we have used the independence of the subtrees and the i.i.d.~nature of the $Z_{n-1}^{(i)}$ variables.

Writing $F_{n-1}(s)$ for the pgf $E\left(s^{Z_{n-1}}\right)$ and summing over all possible values of $k$ (recall that $\widetilde{q}_k$ is the probability that there are $k$ children of the seed particle, i.e., $k$ particles in generation 1) yields
\begin{align}
\widetilde{F}_n(s) = E\left(s^{\widetilde{Z}_n}\right) &= \sum_{k=0}^\infty \widetilde{q}_k \, E\left(\left.s^{\widetilde{Z}_n}\right|\, k\text{ particles in generation }1\right) \nonumber\\
&= \sum_{k=0}^\infty \widetilde{q}_k \left[ E\left( s^{Z_{n-1}}\right)\right]^k \nonumber\\
&= \widetilde{f}\left( F_{n-1}(s) \right) . \label{4}
\end{align}
This equation relates the pgf for $\widetilde{Z}_n$ to the pgf for the subtree quantities $Z_{n-1}$. The next step is to derive an equation that recursively links $Z_{n}$ (the number of particles in a subtree $n$ generations after its birth) to $Z_{n-1}$.
 \begin{figure}
\centering
\epsfig{figure=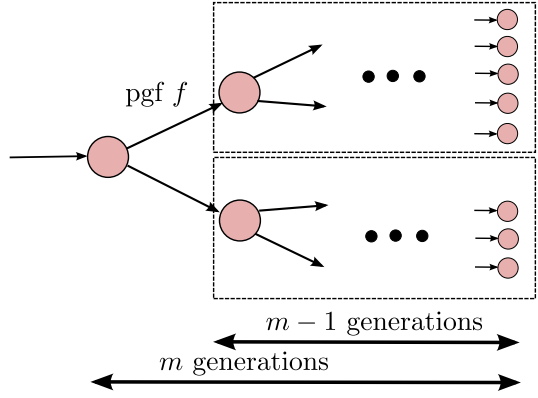,width=7cm}
\caption{Schematic of a subtree generated from a particle that is not a seed; note the number of children of the particle is generated by ${f}$. }\label{schematic_subtree2}
\end{figure}

Figure~\ref{schematic_subtree2} is a schematic view of this relationship. The main subtree in Fig.~\ref{schematic_subtree2} is born with the first particle shown (left of the Figure) and we condition on the number $k$ of children particles of this first particle; recall that $k$ is a random variable with pgf $f(x)$. The number $Z_m$ of particles in the main subtree after $m$ generations is equal to the sum of the $k$ i.i.d.~variables $Z_{m-1}^{(i)}$:
\begin{equation}
Z_m = \sum_{i=1}^k Z_{m-1}^{(i)},
\end{equation}
and so
\begin{align}
E\left( \left. s^{{Z}_m} \right|\, k\text{ particles in generation 1}\right) & = E\left( s^{\sum_{i=1}^k {Z_{m-1}}^{(i)}}\right)\nonumber \\
&= \left[ E\left(s^{Z_{m-1}}\right)\right]^k,\label{5}
\end{align}
as in Eq.~(\ref{4}). Summing over the possible values of $k$ then yields
\begin{align}
F_m(s) = E\left(s^{Z_m}\right) &= \sum_{k=0}^\infty q_k E\left( \left. s^{Z_m}\right| k\text{ particles in generation }1\right)\nonumber\\
&= \sum_{k=0}^\infty q_k \left[ E\left(s^{Z_{m-1}}\right)\right]^k \nonumber\\
&= f\left( F_{m-1}(s)\right). \label{6}
\end{align}

Equation~(\ref{6}) gives a recursion relation for the pgf $F_m(s)$, starting from the initial condition $F_0(s)=s$, corresponding to the tree being seeded from a single particle. Using the result of the recursion Eq.~(\ref{6}) in Eq.~(\ref{4}) then gives the pgf for the number of nodes in generation $n$ of the tree. This characterization of the branching process is called the \emph{backward} approach in \cite{Kimmelbook}, 
 in analogy with the backward Chapman-Kolmogorov equation of Markov processes. An alternative \emph{forward} approach---wherein the states of particles in generation $n+1$ is predicted from the state of the process after $n$ generations---is often used to derive  Eq.~(\ref{6}), but we will find the backward approach easily generalizable to other quantities of interest.

The probability that the tree is terminated at or before generation $n$ is equal to the probability of the tree having zero nodes in generation $n$, which is $\widetilde{F}_n(0)$. The probability that the tree terminates precisely at generation $n$ (i.e., that there are a nonzero number of particles in generation $n$ but each of these has zero offspring) is therefore
\begin{align}
\Omega_n &= \widetilde{F}_n(0)-\widetilde{F}_{n-1}(0) \nonumber\\
&= \widetilde{f}\left(F_{n-1}(0)\right) - \widetilde{f}\left(F_{n-2}(0)\right) \label{7},
\end{align}
where $F_n(0)$ is calculated by iteration from Eq.~(\ref{4}) and the initial condition $F_0(0)=0$. We call $\Omega_n$ the \emph{lifetime distribution of trees}, as it gives the probability that the observed lifetime of a tree is $n$ generations.
\begin{figure}
\centering
\epsfig{figure=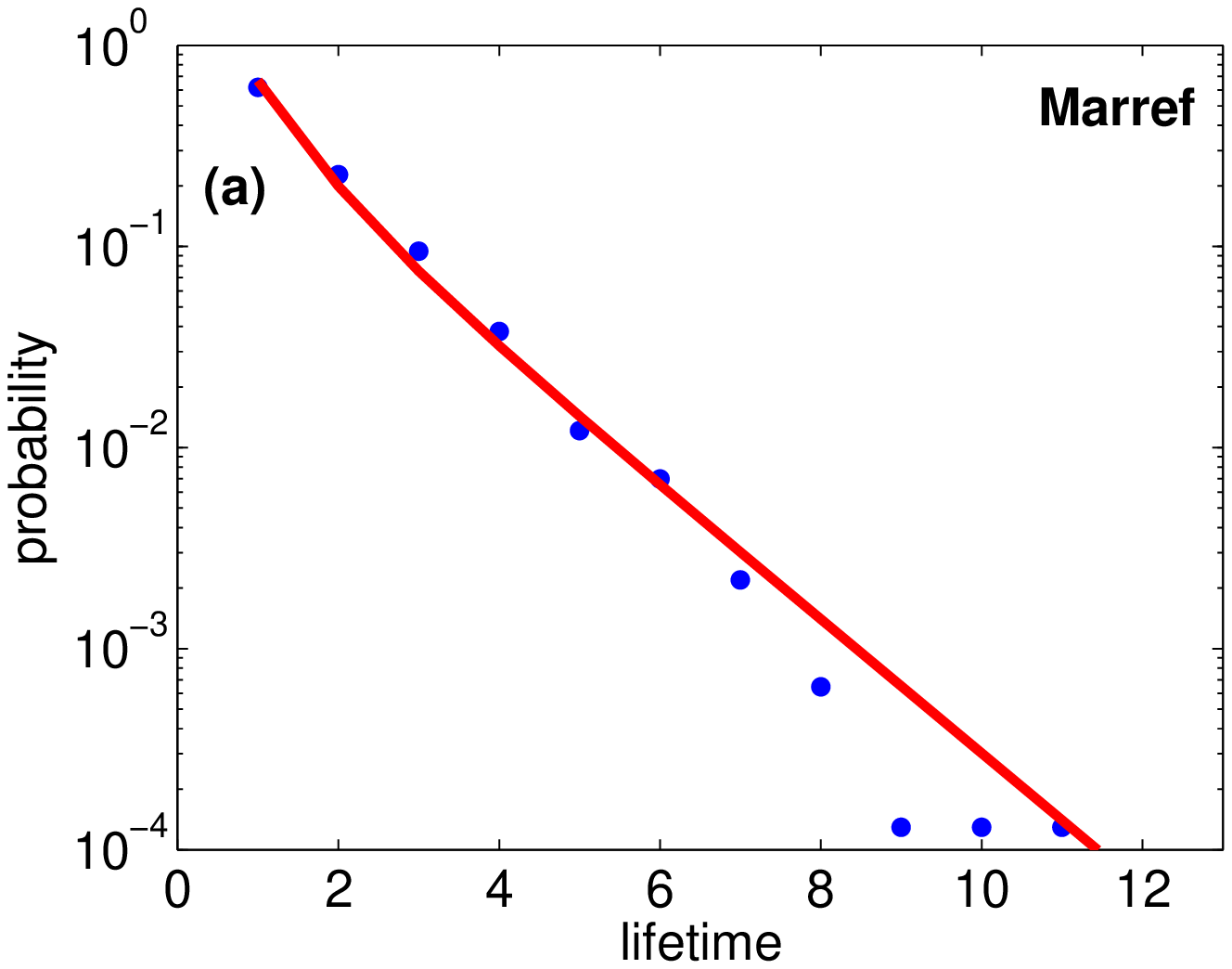,width=8.1 cm}
\epsfig{figure=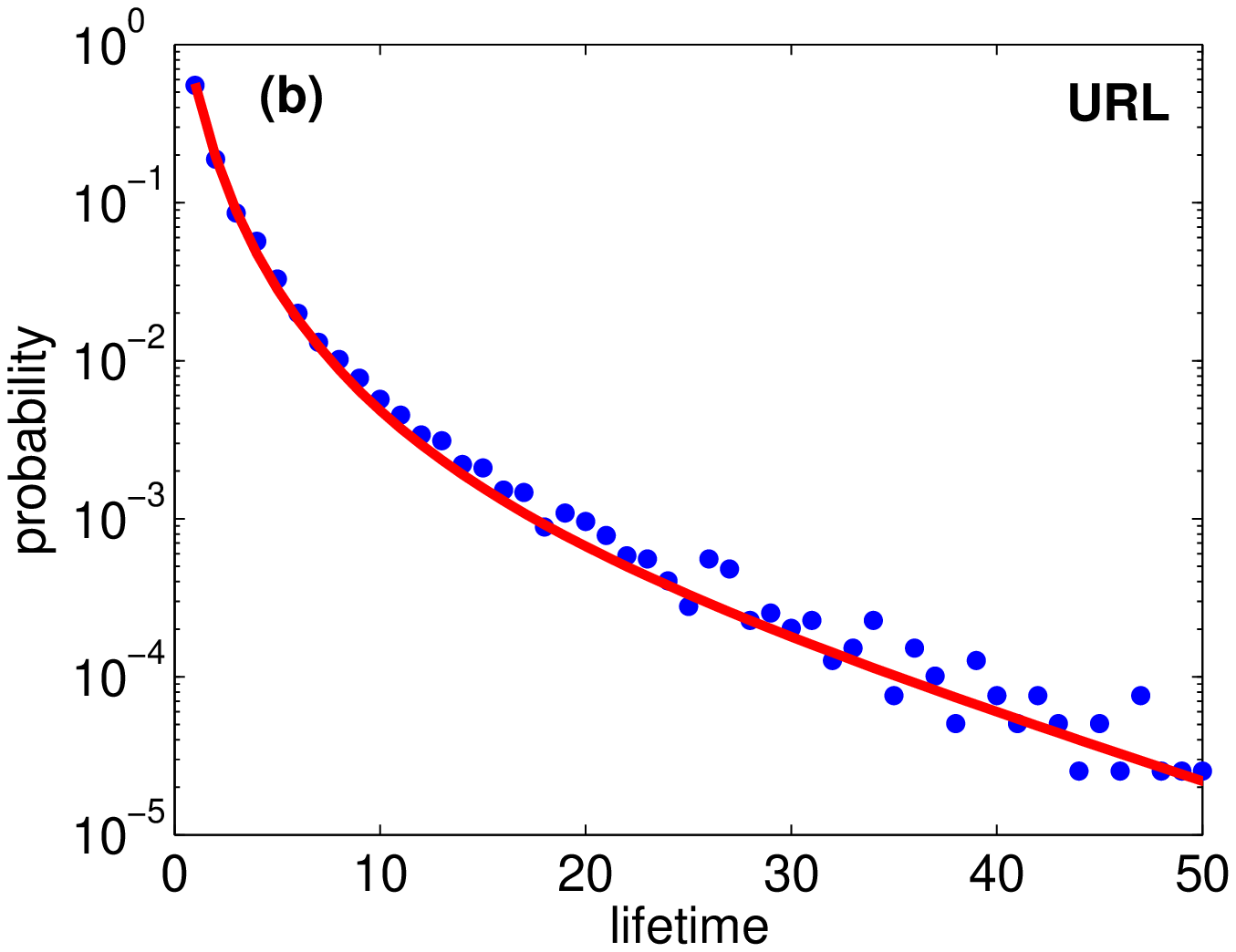,width=8.1 cm} 
\caption{Lifetime distribution of cascades in Marref (left) and URL (right) datasets. Blue symbols are empirical values; red line shows the theoretical distribution from Eq.~(\ref{7}), using the offspring distribution of Eq.~(\ref{3.15}).}\label{fig4}
\end{figure}
See Figure~\ref{fig4} for a comparison of the empirical lifetime distribution with the predictions of Eq.~(\ref{7}), using the offspring distribution fitted in Eq.~(\ref{3.15}).

\subsection{Cascade size}\label{treesize}
A similar approach can be applied to calculate the distribution of tree (cascade) sizes, i.e., the total number of particles that are in all generations of the tree, from the seed at generation 0 up to the last generation of the tree (this quantity is sometimes called the \emph{total progeny} of the tree). We define the random variable $\widetilde{X}_n$ to be the size of the tree observed $n$ generations after its seed particle is born. As before, $\widetilde{X}_n$ can be decomposed into the sum of contributions from each of the subtrees born in generation 1. Conditioning on the seed node having $k$ children particles, we write
\begin{equation}
\widetilde{X}_n = 1+\sum_{i=1}^k X_{n-1}^{(i)}, \label{8}
\end{equation}
where $X_{n-1}^{(i)}$ represents the $i$th i.i.d. subtree size as observed after $n-1$ generations, and the first term counts the seed node of the tree (see Fig.~\ref{schematic_subtree1}). Using identical arguments to those leading to Equations~(\ref{3}) and (\ref{4}), we obtain
\begin{align}
E\left(x^{\widetilde{X}_n}\right) &= E\left( x^{1+\sum_{i=1}^k X_{n-1}^{(i)}}\right)\nonumber\\
&=x \left[ E\left( x^{X_{n-1}}\right)\right]^k  \label{9}
\end{align}
and the pgf $\widetilde{G}_n(x) = E\left( x^{\widetilde{X}_n}\right) = \sum_{j=0}^\infty \text{Prob}\left(\widetilde{X}_n =j\right) x^j$ is then given by
\begin{align}
\widetilde{G}_n(x) &= \sum_{k=0}^\infty \widetilde{q}_k E\left( \left. x^{\widetilde{X}_n}\right| k\text{ particles in generation }1\right) \nonumber\\
&=\sum_{k=0}^\infty \widetilde{q}_k x \left[ E\left(x^{X_{n-1}}\right)\right]^k \nonumber\\
&= x \widetilde{f}\left(G_{n-1}(x)\right), \label{10}
\end{align}
where $G_{n-1}(x) = E\left( x^{X_{n-1}}\right)$ is the pgf for the size of a subtree after $n-1$ generations. Referring now to Fig.~\ref{schematic_subtree2}, the recursion relation for the subtree sizes is derived by first assuming $k$ children in the first generation of the subtree:
\begin{equation}
X_m = 1+ \sum_{i=1}^k X_{m-1}^{(i)},
\end{equation}
and then proceeding as in Equations~(\ref{6}) and (\ref{10}) to obtain the recursion relation
\begin{equation}
G_m(x) = x f\left( G_{m-1}(x)\right), \label{11}
\end{equation}
with initial condition $G_0(x)=x$.

By iterating Eq.~(\ref{11}) for $m=1,2,\ldots,n-1$ and then substituting $G_{n-1}(x)$ into Eq.~(\ref{10}), we obtain the desired pgf $\widetilde{G}_n(x)$ describing the distribution of cascade sizes after $n$ generations. In order to invert the pgf to obtain the distribution of cascade sizes, we iterate Eqs.~(\ref{11}) and (\ref{10}) for a set of $x$ values that are uniformly spaced around the unit circle in the complex $x$-plane, and use a fast Fourier transform to approximate the Cauchy integral
\begin{equation}
\text{Prob}\left(\widetilde{X}_n = j\right) = \left. \frac{1}{j!}\frac{d^j \widetilde{G}_{n}}{d x^j} \right|_{x=0} = \frac{1}{2 \pi i}\oint_C \widetilde{G}_n (x) x^{-(j+1)} dx \label{12},
\end{equation}
as in section S2 of \cite{GleesonPRL14}.

\begin{figure}
\centering
\epsfig{figure=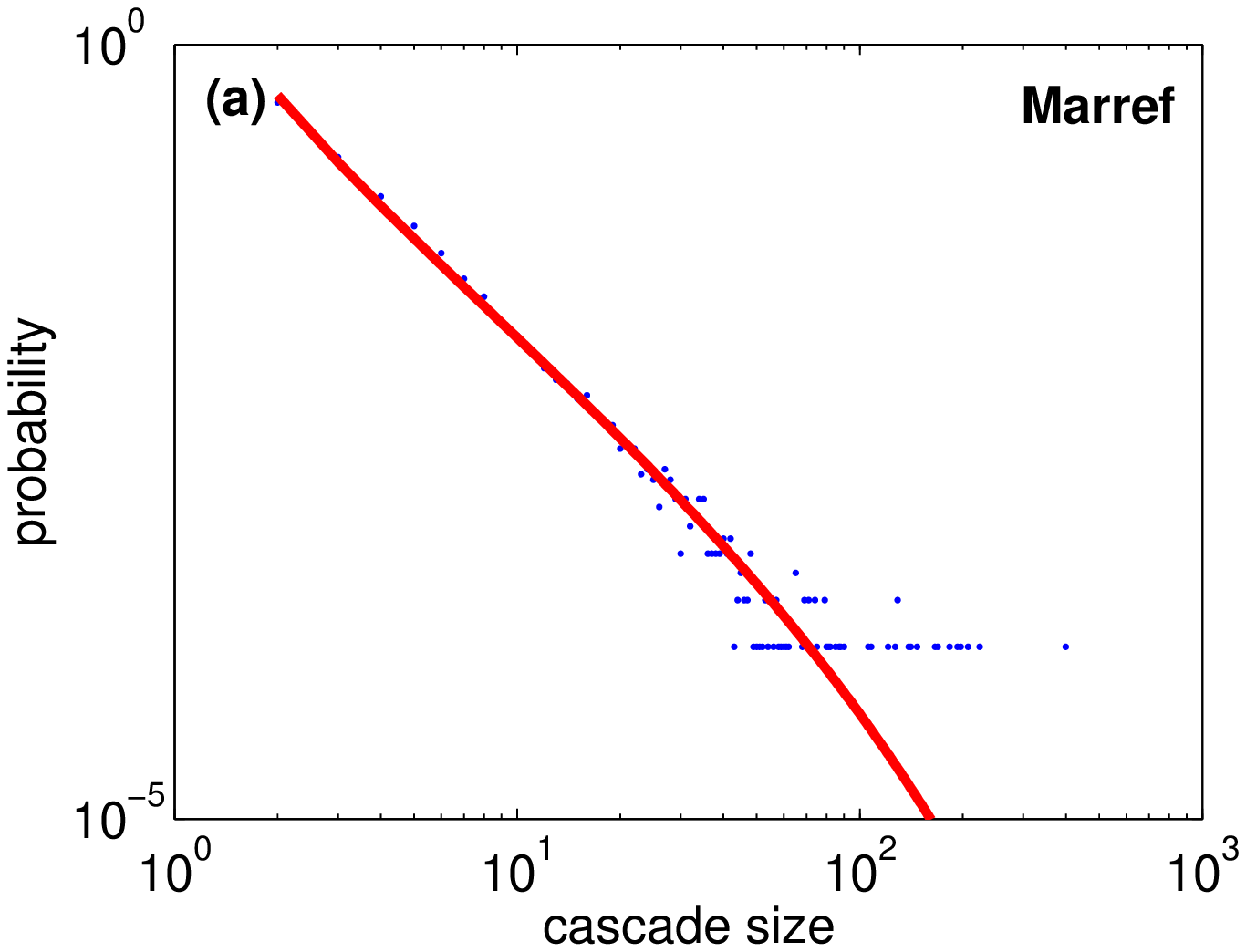,width=8.1 cm}
\epsfig{figure=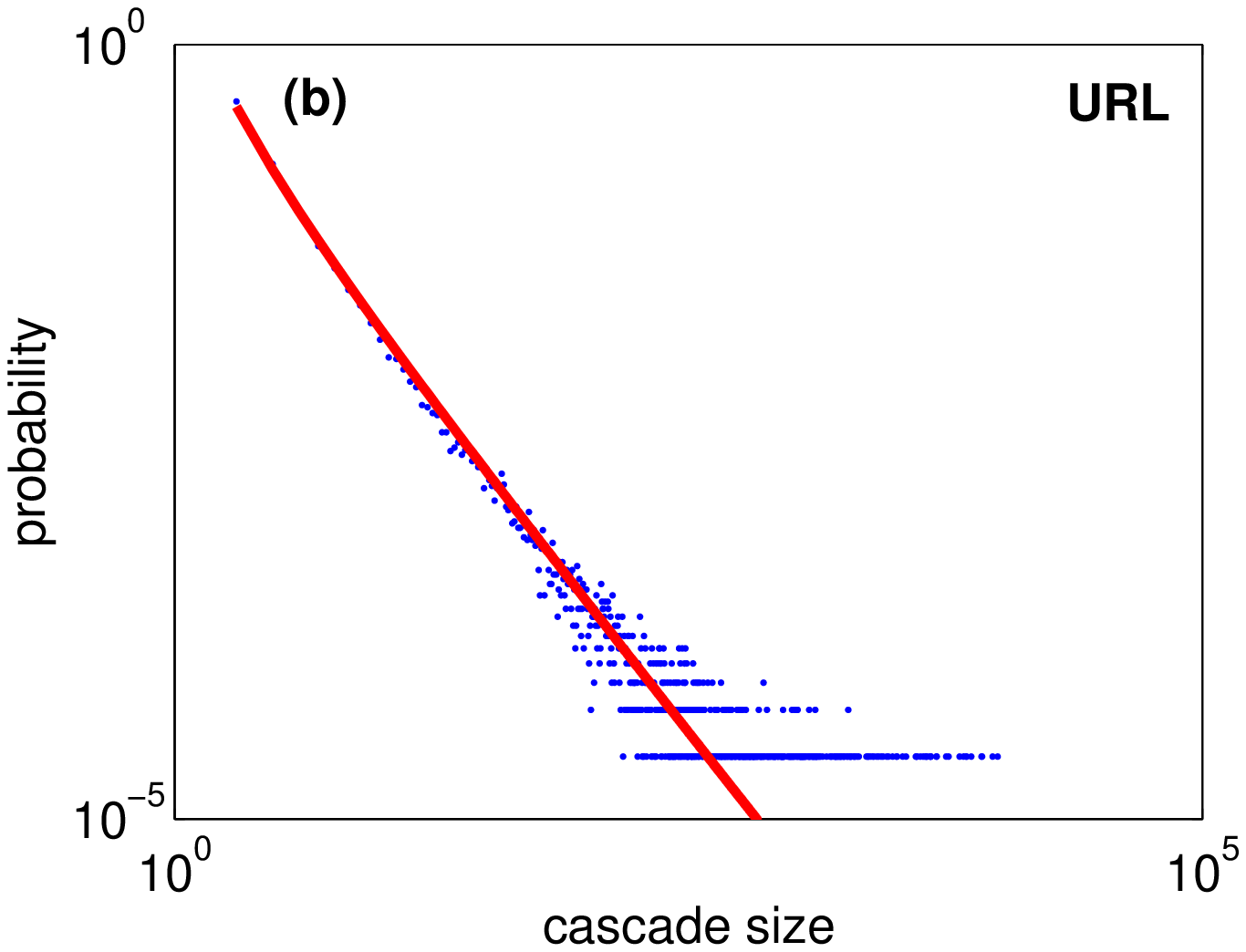,width=8.1 cm}\\ 
\epsfig{figure=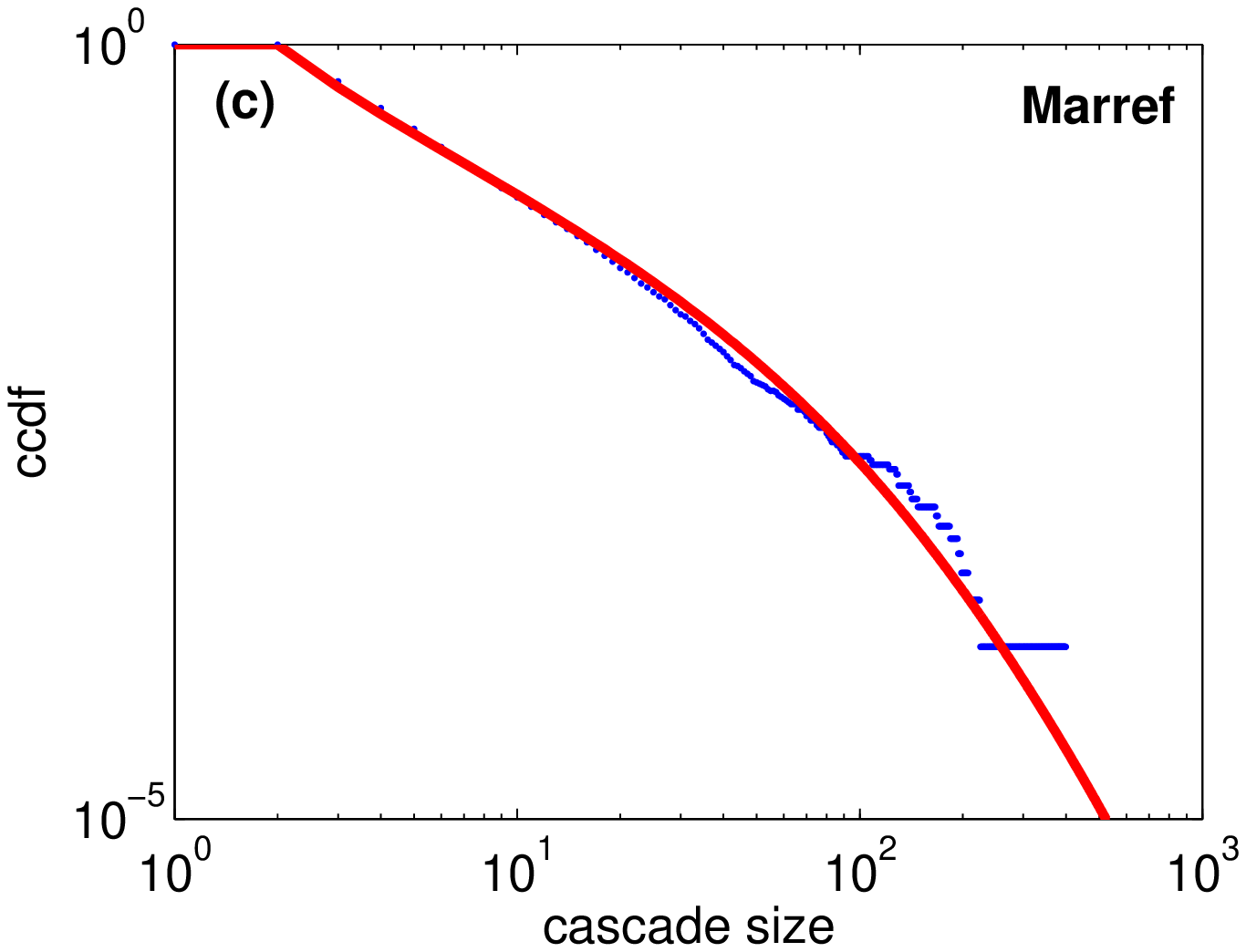,width=8.1 cm}
\epsfig{figure=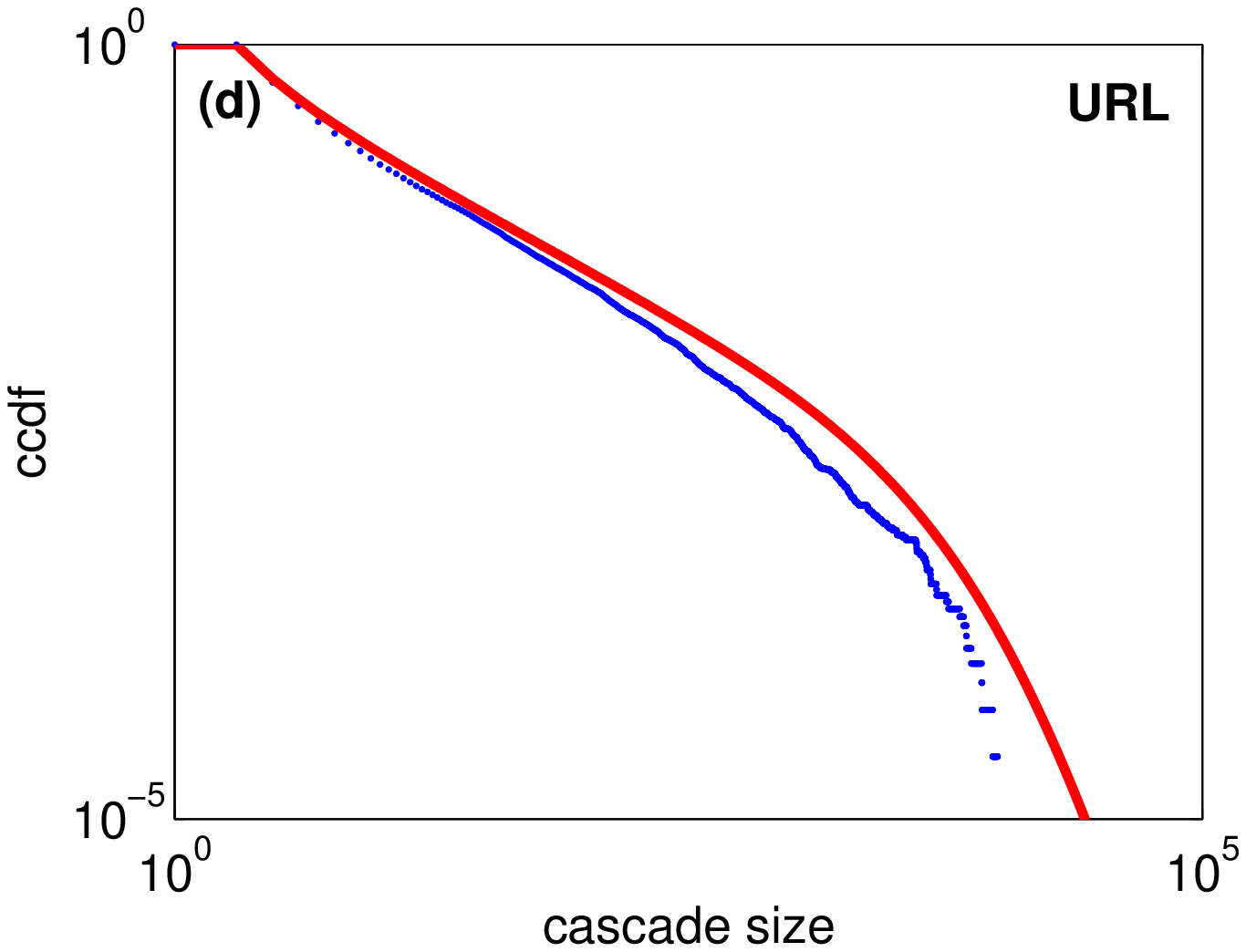,width=8.1 cm}
\caption{Cascade size distributions: pdfs (top panels) and ccdfs (bottom panels) for Marref (left) and URL (right). Blue symbols are empirical values; red line shows the theoretical distribution from Sec.~\ref{treesize}, using the offspring distribution of Eq.~(\ref{3.15}). }\label{fig3}
\end{figure}

Figure~\ref{fig3} shows the large-$n$ limit of the cascade size distribution, and compares it with the empirical distribution. The good agreement between this theoretical prediction and the empirical results gives further support to the usage of branching process descriptions for such data.

\subsection{Average tree depth and structural virality} \label{sec4.3}
In this subsection, we build on the approach used in Sec.~\ref{sec4.1} to derive results for measures of the shape of cascade trees, which are of considerable interest in analyses of Twitter \cite{Goel15,OSullivan17}. We focus on the distribution (and expected value) of two quantities \cite{Goel15}: the average depth of a tree, and the structural virality of a tree.
\subsubsection{Average tree depth}
\begin{figure}
\centering
\epsfig{figure=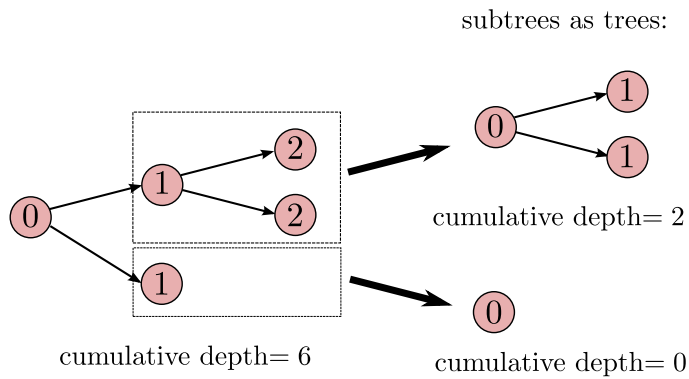,width=9cm}
\caption{This is a tree of size 5. Each of the 5 particles is labelled by its depth (its generation plus 1). The cumulative depth of the tree is 0+1+1+2+2=6, and so the average depth of this tree is $6/5$. Note that the cumulative depth of the top subtree is 0+1+1=2, also that each node in the subtree has a depth that is one larger than its value when considered as part of the main tree: See Eq.~(\ref{14}).}\label{figschematicavgdepth}
\end{figure}

To calculate the \emph{average depth} of a sample tree, we first sum the depths (generation numbers) of all particles in the tree to obtain the \emph{cumulative depth} of the tree, and then divide this by the size of the tree (the total number of particles in the tree), see Fig.~\ref{figschematicavgdepth}.
In this subsection we generalize the methods used in Sec.~\ref{treesize} to calculate the joint distribution of tree size and cumulative depth, and hence to find a formula for the \emph{expected average tree depth} (EATD). In the ensemble of trees generated by the branching process, each tree has its own average depth, and the EATD is the mean of the average depths over all trees in the ensemble. We believe the formula we derive (Eq.~(\ref{23})) is novel.


We extend the approach of Sec.~\ref{treesize} to consider the joint distribution of $\widetilde{X}_n$ (the tree size after $n$ generations) and of $\widetilde{Y}_n$, which is the random variable giving the cumulative depth of the tree after $n$ generations. We define the two-variable pgf $\widetilde{H}_n(x,y)$ as
\begin{equation}
\widetilde{H}_n(x,y) = E\left( x^{\widetilde{X}_n} y^{\widetilde{Y}_n}\right) = \sum_{j, \ell = 0}^\infty \text{Prob}\left(\widetilde{X}_n=j\text{ and }\widetilde{Y}_n = \ell\right)x^j y^\ell. \label{13}
\end{equation}
As in earlier sections, we relate the variables $\widetilde{X}_n$ and $\widetilde{Y}_n$ to subtree quantities, and begin by assuming that the seed node (in Fig.~\ref{schematic_subtree1} for example) has $k$ children. Each of the $k$ children generates a subtree with (after $n-1$ generations of the subtree) respective random-variable pairs $\left(\widetilde{X}_{n-1}^{(i)}, \widetilde{Y}_{n-1}^{(i)}\right)$ for $i=1,2,\ldots,k$.

The relationship between $\widetilde{X}_n$ and $X_{n-1}^{(i)}$ is given by Eq.~(\ref{8}) but we must now also find an analogous expression for $\widetilde{Y}_n$. We define $Y_{n-1}^{(i)}$ to be the cumulative depth of the $i$th i.i.d. subtree. Notice (see Fig.~\ref{figschematicavgdepth}) that when we add the $Y_{n-1}^{(i)}$ values for all the subtrees, each node of the subtree has a depth that is one less that its depth in the main tree. Therefore, the $i$th subtree contributes to $\widetilde{Y}_n$ a total of $Y_{n-1}^{(i)} + X_{n-1}^{(i)}$, where the second term adds one for each node in the subtree. From this relationship, we obtain
\begin{equation}
\widetilde{Y}_n = \sum_{i=1}^k\left(Y_{n-1}^{(i)} + X_{n-1}^{(i)}\right), \label{14}
\end{equation}
and with Eq.~(\ref{8}) we find the pgf relations
\begin{align}
\widetilde{H}_n(x,y) & = \sum_{k=0}^\infty \widetilde{q}_k E\left(\left. x^{\widetilde{X}_n} y^{\widetilde{Y}_n}\right|k\text{ particles in generation }1\right) \nonumber\\
&= \sum_{k=0}^\infty \widetilde{q}_k E\left(x^{1+\sum_{i=1}^k X_{n-1}^{(i)}} y^{\sum_{i=1}^k \left(Y_{n-1}^{(i)}+X_{n-1}^{(i)}\right)}\right) \nonumber\\
& = x \sum_{k=0}^\infty \widetilde{q}_k E\left( (x y)^{\sum_{i=1}^k X_{n-1}^{(i)}}  y^{\sum_{i=1}^k Y_{n-1}^{(i)}}\right) \nonumber\\
& = x \sum_{k=0}^\infty \widetilde{q}_k E\left( (x y)^{X_{n-1}} y^{Y_{n-1}}\right) \nonumber\\
& = x \widetilde{f}\left( H_{n-1}(x y, y)\right)\label{15},
\end{align}
where $H_{n-1}(x,y)= E\left(x^{X_{n-1}} y^{Y_{n-1}}\right)$.

Addressing the recursion relation for the subtrees in a similar fashion leads (as in Sec.~\ref{treesize}) to
\begin{equation}
H_{m}(x,y) = x f \left( H_{m-1}(x y ,y )\right), \label{16}
\end{equation}
with initial condition $H_0(x,y)=x$ (since a single particle is  a tree of size 1, with zero depth). Iterating Eq.~(\ref{16}) for $m=1,2,\ldots,n-1$ and substituting into Eq.~(\ref{15}) yields the pdf $\widetilde{H}_n(x,y)$ for the joint distribution of trees size and cumulative depth after $n$ generations.

We can use this joint distribution to calculate the EATD for trees of $n$ generations as
\begin{align}
d_n &= \sum_{j=0}^\infty \sum_{\ell=0}^\infty \text{Prob}\left(\widetilde{X}_n=j\text{ and }\widetilde{Y}_n =\ell\right) \frac{\ell}{j}\nonumber\\
&= \int_0^1 \frac{1}{x}\left.\frac{\partial \widetilde{H}_n}{\partial y}\right|_{y=1} d x, \label{17}
\end{align}
as can be verified by term-by-term differentiation and integration of the series in Eq.~(\ref{13}). Taking the $n\to \infty$ limit in order to include all trees, Eq.~(\ref{16}) give a self-consistent equation for $H_\infty(x,y)$ and it can be differentiated with respect to $x$ to yield
\begin{equation}
\left. \frac{\partial H}{\partial x }\right|_{y=1} = \frac{ f\left(H(x,1)\right)}{1- x f'\left(H(x,1)\right)}, \label{18}
\end{equation}
where, for simplicity, we drop the subscript from $H_\infty$ for the remainder of this section.
Similarly, differentiation of Eq.~(\ref{16}) with respect to $y$ gives
\begin{equation}
\left. \frac{\partial H}{\partial y}\right|_{y=1} = x f'\left(H(x,1)\right) \left[ \left. x \frac{\partial H}{\partial x}\right|_{y=1} +\left.  \frac{\partial H}{\partial y}\right|_{y=1} \right],
\end{equation}
which can be solved for $\left. \frac{\partial H}{\partial y}\right|_{y=1}$, after substituting for  $\left. \frac{\partial H}{\partial x}\right|_{y=1}$ from Eq.~(\ref{18}):
\begin{equation}
 \left.\frac{\partial H}{\partial y}\right|_{y=1} = \frac{ x^2 f'\left(H(x,1)\right) f\left(H(x,1)\right) }{\left[1-x f'\left(H(x,1)\right)\right]^2}. \label{19}
 \end{equation}
Differentiating the $n\to\infty$ limit of Eq.~(\ref{15}) with respect to $y$ yields
\begin{equation}
\left. \frac{\partial \widetilde H}{\partial y}\right|_{y=1} = x \widetilde{f}'\left(H(x,1)\right) \left[ \left. x \frac{\partial H}{\partial x}\right|_{y=1} +\left.  \frac{\partial H}{\partial y}\right|_{y=1} \right], \label{20}
\end{equation}
and substituting from Eqs.~(\ref{18}) and (\ref{19}) gives
\begin{equation}
 \left.\frac{\partial \widetilde H}{\partial y}\right|_{y=1} = \frac{ x^2 \widetilde{f}'\left(H(x,1)\right) f\left(H(x,1)\right) }{\left[1-x f'\left(H(x,1)\right)\right]^2}. \label{21}
 \end{equation}
Thus, the expected  average tree depth over all trees is given by Eq.~(\ref{17}) as
\begin{equation}
d = \int_0^1  \frac{ x \widetilde{f}'\left(H(x,1)\right) f\left(H(x,1)\right) }{\left[1-x f'\left(H(x,1)\right)\right]^2} d x.\label{22}
\end{equation}
Noting that Eq.~(\ref{16}) relates $H(x,1)$ to $x$ through the implicit relation
\begin{equation}
H(x,1) = x f\left( H(x,1) \right), \label{22b}
\end{equation}
we make the change of integration variable $x\mapsto h$ defined implicitly by $x=h/f(h)$ (with $dx = \left( f(h)- h f'(h)\right)/f^2(h) d h$), to yield a simple integral formula for the EATD:
\begin{equation}
d = \int_0^1 \frac{h \widetilde{f}'(h)}{f(h) - h f'(h)} d h. \label{23}
\end{equation}
 This remarkably simple formula is easily evaluated once the offspring distributions $f$ and $\widetilde f$ of the branching process are given.
  In Table~\ref{tab2} we show that it agrees with Monte Carlo simulations and also gives quite a reasonably accurate estimate of the values found from the empirical data.
\begin{figure}
\centering
\epsfig{figure=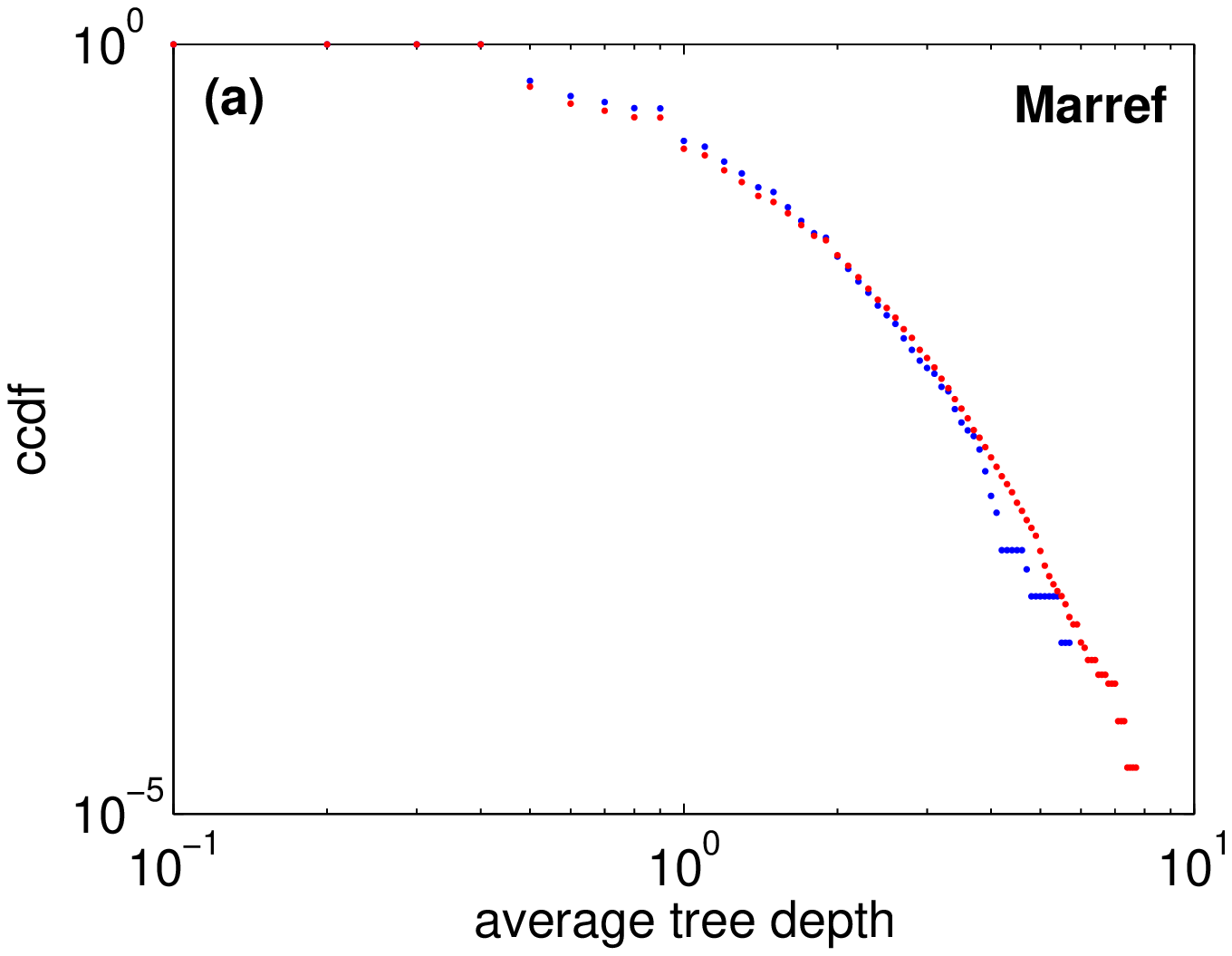,width=8.1 cm}
\epsfig{figure=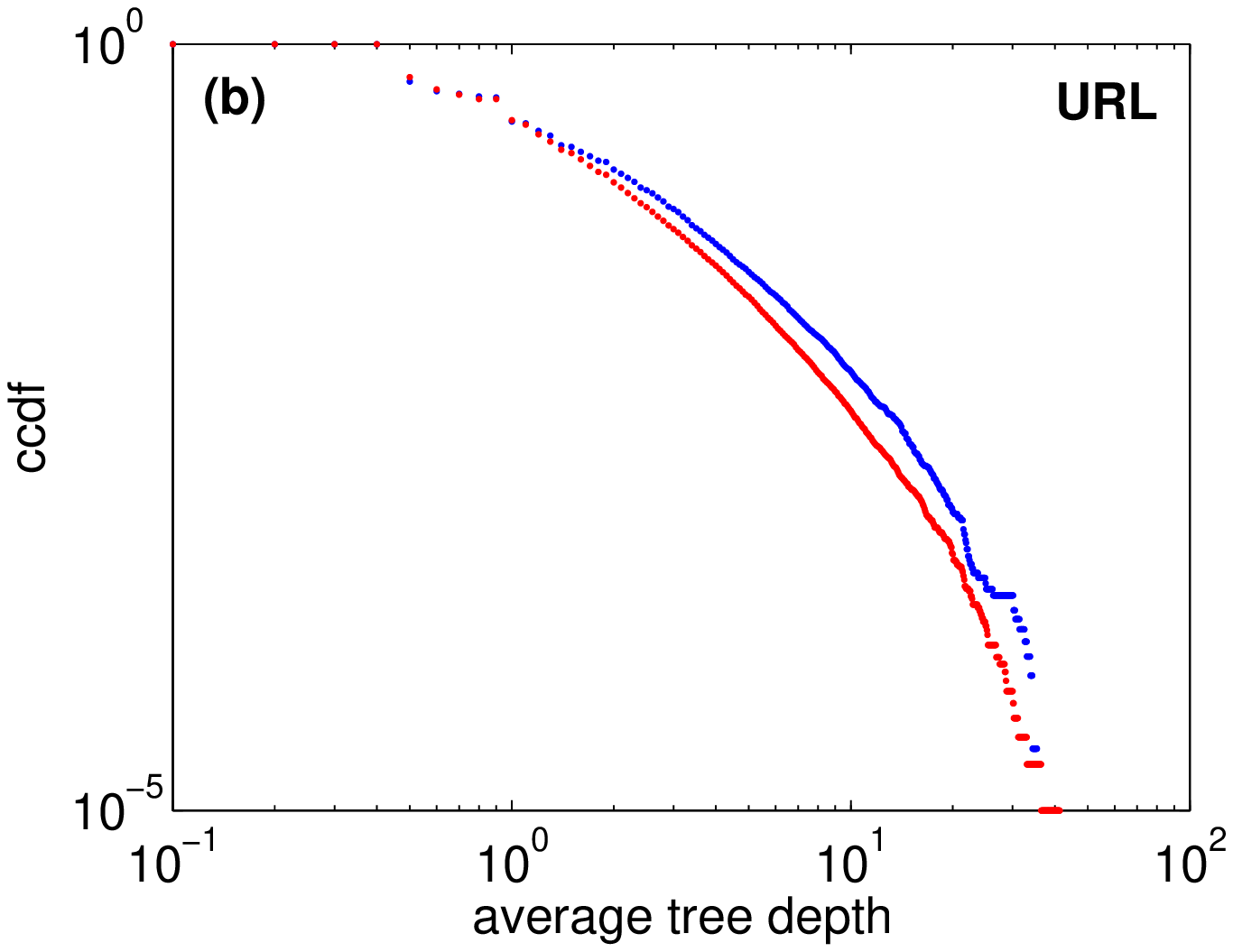,width=8.1 cm}\\ 
\epsfig{figure=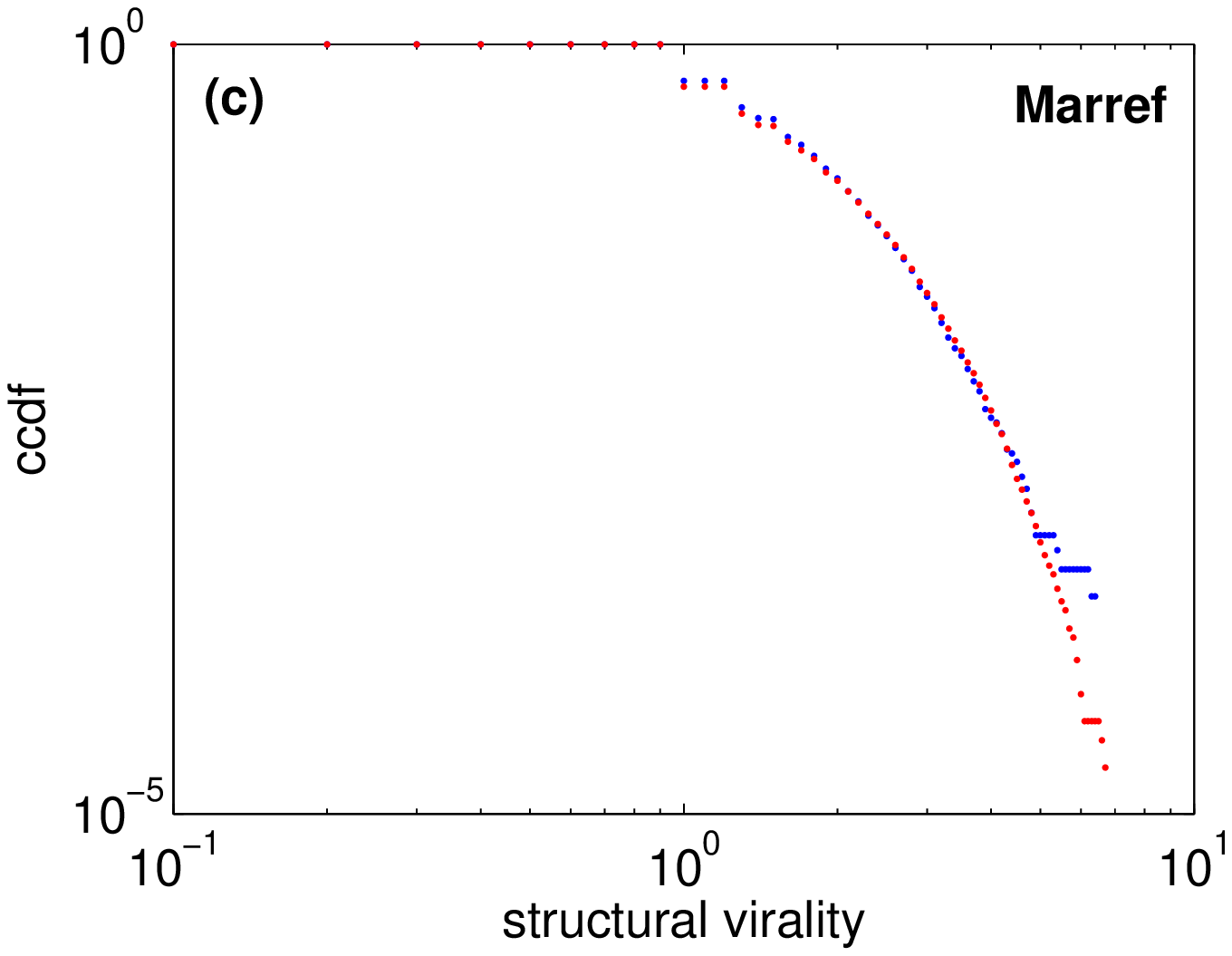,width=8.1 cm}
\epsfig{figure=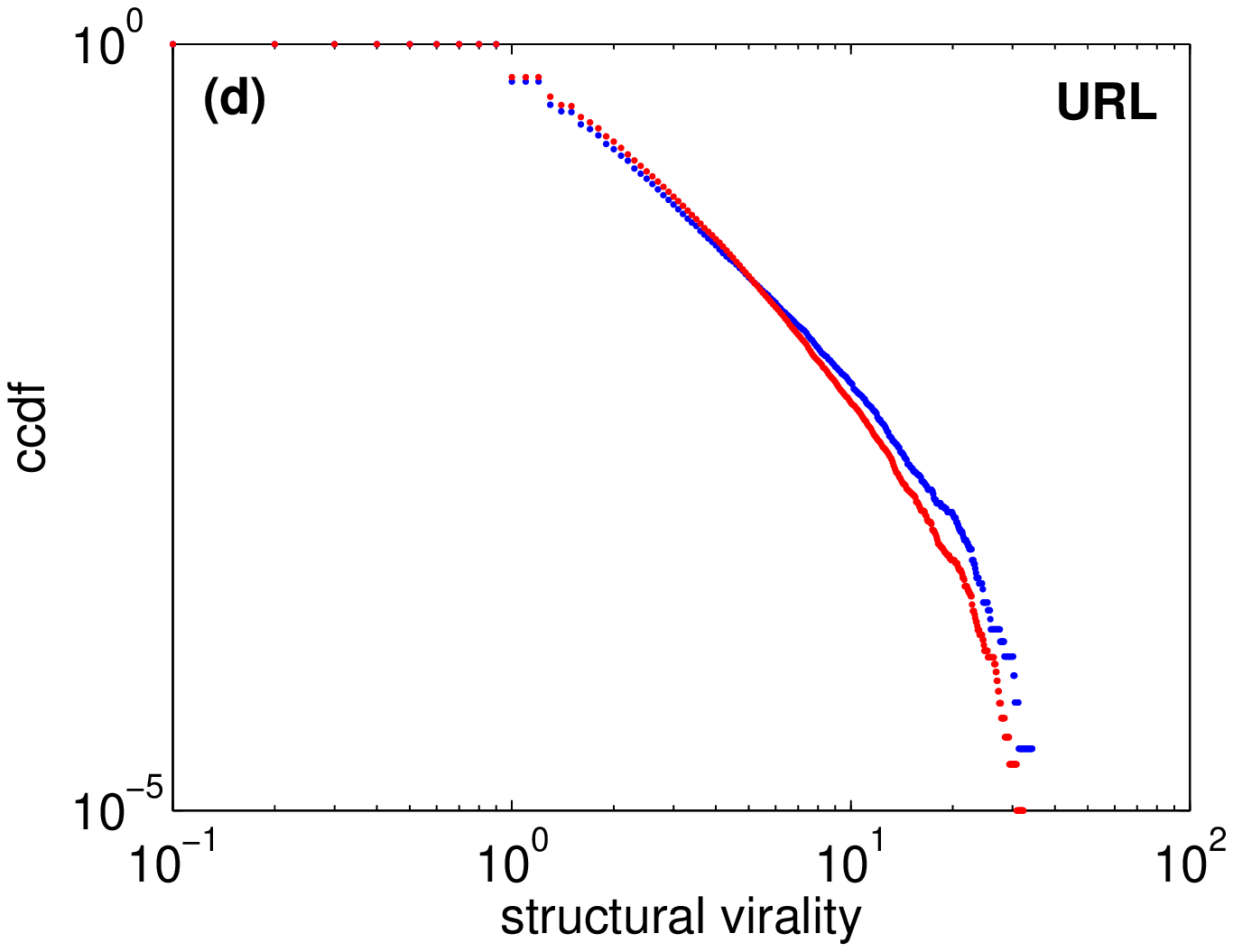,width=8.1 cm}
\caption{Ccdfs of average tree depth (top panels) and structural virality (bottom panels) for Marref (left) and URL (right). Blue symbols are empirical distributions; red symbols are from Monte Carlo simulations of branching processes with offspring distribution $q_\ell$ given by Eq.~(\ref{3.15}), and by the corresponding distribution $\widetilde{q}_\ell$ for the seed's offspring.
}\label{fig5}
\end{figure}
\begin{table}
\begin{center}
\begin{tabular}{|c||c|c|}
\hline
    & Marref & URL  \\
\hline\hline 
integral & 0.862 & 1.22  \\ \hline
Monte Carlo & 0.862\, (0.859, 0.866) & 1.22\, (1.21, 1.23) \\ \hline
data & 0.899\, (0.887, 0.912) & 1.34\, (1.32, 1.36) \\ \hline
 \hline
\end{tabular}
\end{center}
\caption{Expected average tree depth (EATD)
from the integral formula of Eq.~(\ref{23}), compared with Monte Carlo simulations ($10^5$ realizations) and the data values.
Bootstrap intervals given for the latter two cases show quantile 0.025 to quantile 0.975 (i.e., $95\%$ of cases) for the expected average tree depth, using $10^3$ bootstrap samples.
} 
\label{tab2}
\end{table}

\subsubsection{Structural virality}
\begin{table}
\begin{center}
\begin{tabular}{|c||c|c|} 
\hline
    & Marref & URL  \\
\hline\hline
integral & 1.44 & 1.81  \\ \hline
Monte Carlo & 1.440\, (1.436, 1.443) & 1.82\, (1.81, 1.83) \\ \hline
data & 1.47\, (1.46, 1.49) & 1.77\, (1.75, 1.78) \\ \hline
 \hline
\end{tabular}
\end{center}
\caption{Expected structural virality from the integral formula of Eq.~(\ref{SVintegral}), compared with Monte Carlo simulations ($10^5$ realizations) and the data values.
Bootstrap intervals given for the latter two cases show quantile 0.025 to quantile 0.975  (i.e., $95\%$ of cases) for the expected structural virality, using $10^3$ bootstrap samples.}
\label{tab3}
\end{table}


The \emph{structural virality} of a tree with size $n>1$ was introduced in Goel et al.~\cite{Goel15} as
\begin{equation}
\frac{1}{n(n-1)} \sum_{i=1}^n \sum_{j=1}^n d_{i j},
\end{equation}
where $d_{i j}$ is the graph distance from node $i$ to node $j$. The distribution of this metric across an ensemble of trees was used to fit models to data in \cite{Goel15}.

As noted in \cite{Goel15}, the structural virality of a tree is closely related to its \emph{Wiener index}, defined by $\sum_{i=1}^n \sum_{j=1}^n d_{i j}$.
If we consider the expected value of the Wiener index across the ensemble of trees generated by the branching process, we can usefully adapt the approach of Entringer et al.~\cite{Entringer94}, with the aim of calculating the expected  structural virality for the exnsemble. Entringer et al.~define a generating function $W(x)$ so that the coefficient of $x^n$ in the power series is the contribution of trees of size $n$ to the ensemble-averaged Wiener index (note that $W(x)$ is not a probability generating function). Their Eq.~(3.5) is
\begin{equation}
W(x) = D(x) + x f'(G)W(x)+x f''(G)\left[D(x)+x G'\right]x G', \label{Ent1}
\end{equation}
where $D(x)$ is our $\left.\frac{\partial H}{\partial y}\right|_{y=1}$ from Eq.~(\ref{19}) and $G(x)=H(x,1)$ is the $m\to\infty$ cascade size pgf from Eq.~(\ref{11}). The first term in Eq.~(\ref{Ent1}) comes from considering pairs of vertices $u$ and $v$ where one of $u$ or $v$ is the root of the tree. The second term arises from the case where $u$ and $v$ belong to the same subtree, and the third term stems from $u$ and $v$ belonging to different subtrees, see Sec.~3 of \cite{Entringer94} for details.

We extend the approach of Eq.~(\ref{Ent1}) to the case where the seed node of the tree has offspring distribution with pgf $\widetilde{f}$, to get an analogous equation for $\widetilde{W}(x)$:
\begin{equation}
\widetilde{W}(x) = \widetilde{D}(x) + x \widetilde{f}'(G)W(x)+x \widetilde{f}''(G)\left[D(x)+x G'\right]x G', \label{Ent2}
\end{equation}
where $\widetilde{D}(x)$ is $\left.\frac{\partial \widetilde{H}}{\partial y}\right|_{y=1}$ as given by Eq.~(\ref{21}).

Solving Eq.~(\ref{Ent1}) for $W(x)$ and substituting into Eq.~(\ref{Ent2}) enables us to determine $\widetilde{W}(x)$. The expected structural virality for the ensemble of trees is then given by
\begin{equation}
s = \sum_{n=2}^\infty\frac{\widetilde{W}_n}{n(n-1)}, \label{Ent3}
\end{equation}
where $\widetilde{W}_n$ is the coefficient of $x^n$ in the power series of $\widetilde{W}(x)$. The value of $s$ can be calculated from the generating function $\widetilde{W}(x)$ by a double integration:
\begin{align}
s & = \int_0^1 \left(\int_0^y \frac{\widetilde{W}(x)}{x^2} dx\right) dy \nonumber\\
 & = \int_0^1 \frac{\widetilde{W}(x)}{x^2}(1-x)\, dx, \label{Ent4}
 \end{align}
 where the second equation follows from changing the order of integration, i.e., using the identity
 \begin{equation}
 \int_0^1 \int_0^y (\cdot) \, dx\, dy = \int_0^1 \int_x^1 (\cdot) \, dy\, dx.
 \end{equation}

Combining these results and then making the same change of variable as for Eq.~(\ref{23}) yields an integral formula for the expected structural virality:
\begin{equation}
s=\int_0^1 \frac{f\left(f-h\right)\left(f \tilde{f}' + h f \tilde{f}''-h f' \tilde{f}' + h^2 \left(f'' \tilde{f}'-\tilde{f}'' f'\right)\right)}{\left(f-h f'\right)^3}\, dh \label{SVintegral}
\end{equation}

Table~\ref{tab3} shows that this formula agrees with Monte Carlo simulations of the branching process, and also matches reasonably well to the average structural virality of the ensemble of empirical trees in both datasets.

\subsubsection{Asymptotic analysis}
The integral formulas derived for the expected average tree depth (Eq.~(\ref{23})) and the expected structural virality (Eq.~(\ref{SVintegral})) enable us to analytically study the impact of the spreading process upon these measures. Such understanding can assist in the fitting of information-spreading models to empirical data. In Figure 2 of Ref.~\cite{Goel15}, for example, large-scale numerical simulations are used to calculate the dependence of the expected structural virality on the branching number, and this information is then used to guide model parameter fitting.

We are therefore motivated to examine how the integrals in Eqs.~(\ref{23}) and (\ref{SVintegral}) depend upon the form of the offspring distribution (through its pgf $f(x)$) and in particular on the branching number $\xi = f'(1)$. For simplicity we will restrict ourselves in this section to the case where $\widetilde{f}(x) = f(x)$, i.e., assuming that the seed node's offspring distribution is the same as that of the later generations.

\begin{figure}
\centering
\epsfig{figure=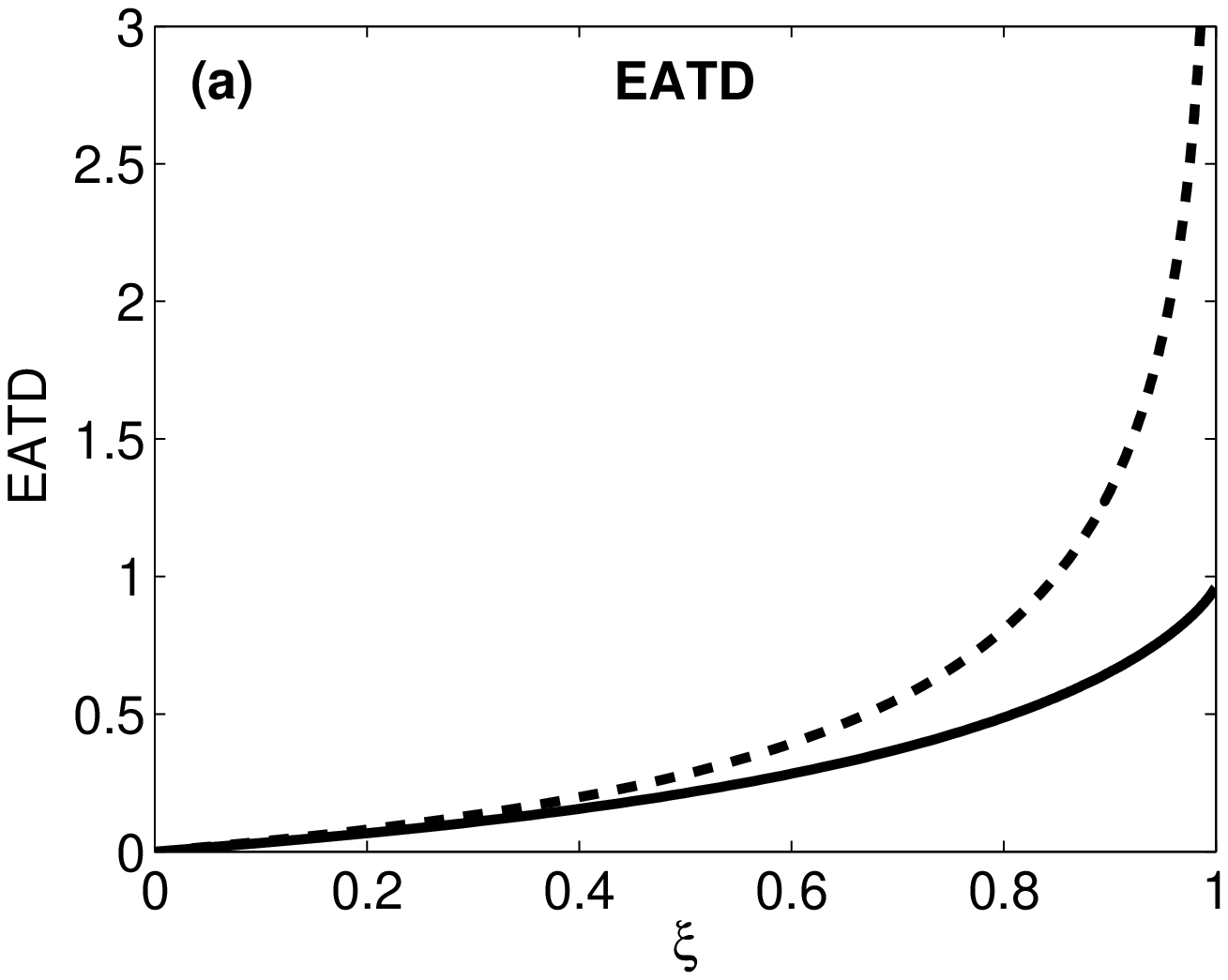,width=8.1 cm}
\epsfig{figure=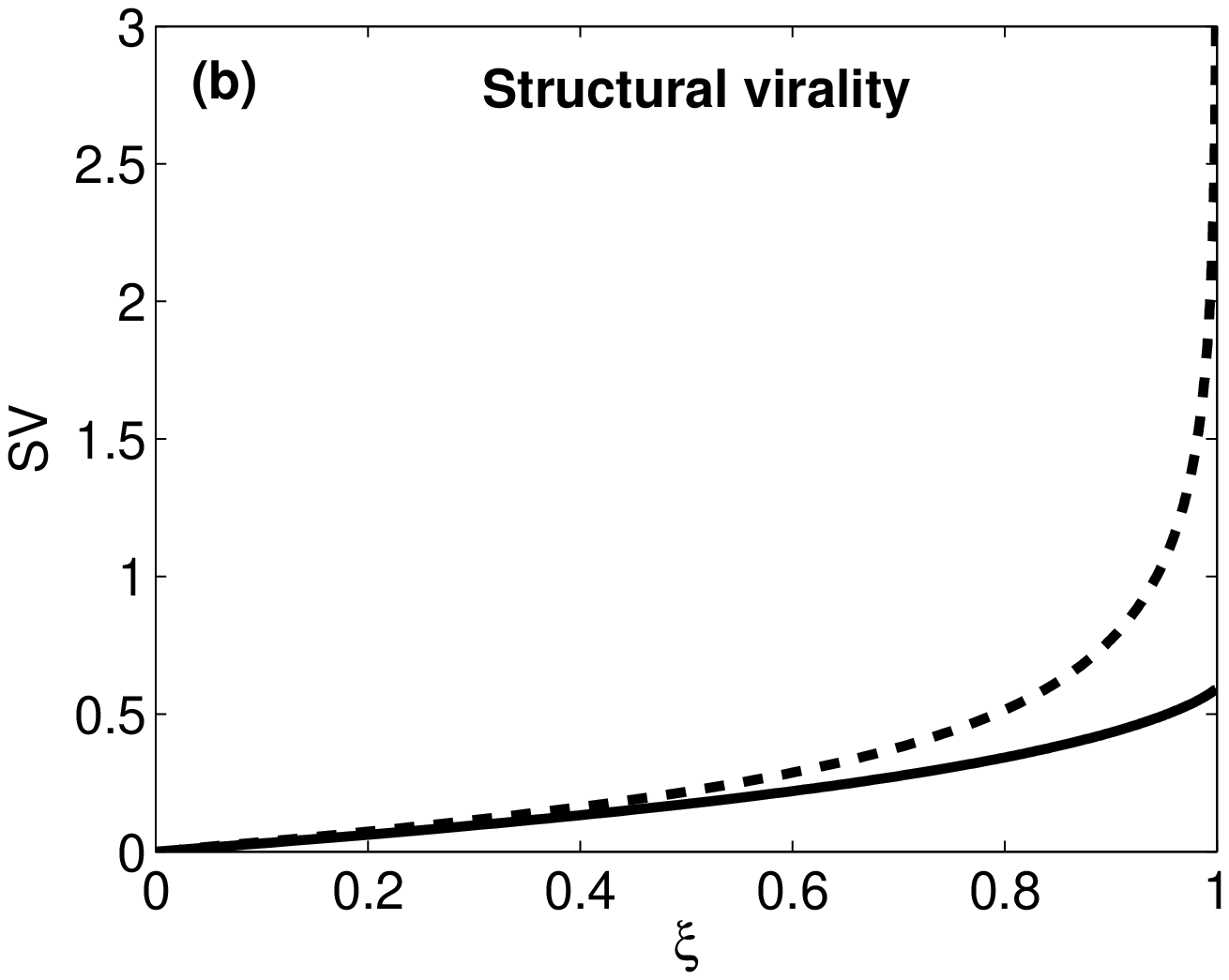,width=8.1 cm}  
\caption{Results of the integral formulas in Eqs.~(\ref{23}) and (\ref{SVintegral}) for Expected Average Tree Depth (left) and Structural Virality (right) for trees with binary fission offspring distribution (dashed curves) and with power-law (tail exponent $\gamma=2.5$) offspring distribution (solid curves). }\label{figasympt}
\end{figure}

First we note that both integrals may be performed exactly in the special case of a binary fission process \cite{AthreyaNeybook}, where each parent has either zero or two children:
\begin{equation}
f_\text{bf}(x) = \left(1-\frac{\xi}{2}\right) + \frac{\xi}{2} x^2.
\end{equation}
The exact integrals for EATD and expected structural virality in this case are
\begin{align}
d_{\text{bf}} &= 2\sqrt{\frac{2-\xi}{\xi}} \text{ArcTanh}\sqrt{\frac{\xi}{2-\xi}} - 2 \nonumber\\
s_{\text{bf}} &= 1 - \frac{\xi}{2} + (2-\xi)\log\left(\frac{2-\xi}{2-2\xi}\right) - \sqrt{\frac{2-\xi}{\xi}}\text{ArcTanh}\sqrt{\frac{\xi}{2-\xi}}
\end{align}
and each shows a logarithmic divergence as the branching number $\xi$ approaches the critical value of 1 from below (see dashed curves in Fig.~\ref{figasympt}).

In fact, this logarithmic divergence as $\xi\to 1$ is not unique to the exactly-solvable binary fission example. Indeed, asymptotic analysis of the integrals shows that a similar divergence occurs for any offspring distribution that has a finite value of $f''(1)$, meaning that the second moment of the offspring distribution is finite. The integrands in Eqs.~(\ref{23}) and (\ref{SVintegral}) are singular at $h=1$, and the form of the singularity can be understood using the expansion of $f(h)$ about $h=1$:
\begin{align}
f(h)& \sim f(1) + f'(1)(h-1)+\frac{1}{2}f''(1) (h-1)^2 + \ldots \nonumber\\
    & = 1- \xi (1-h) + \frac{1}{2}f''(1) (1-h)^2 + \ldots \,\text{ as } h\to 1^-.
\end{align}
The integrand of Eq.~(\ref{23}), for example, has leading-order expansion
\begin{equation}
  \sim \frac{\xi}{1-\xi+f''(1)(1-h)}\, \text{ as } h\to 1^-
\end{equation}
and so the integral diverges logarithmically as $\xi\to 1$; the same asymptotic behaviour is found for the integrand in Eq.~(\ref{SVintegral}). Hence the behaviour of the dashed curves in Fig.~\ref{figasympt} is quite generic for offspring distributions with finite second moments.

Offspring distributions with infinite second moments are also of interest, as they relate to heavy-tailed follower distributions in the  Twitter network \cite{Kwak10,Gleeson16}. An important example is the case of a power-law tail, i.e.,
\begin{equation}
q_\ell \sim D\, \ell^{-\gamma} \, \text{ as } \ell \to \infty,
\end{equation}
for constant $D$ and for values of the exponent between 2 and 3. The asymptotic series for $f(h)$ as $h\to 1^-$ is given in this case by \cite{Wilfbook,GleesonPRL14}
\begin{equation}
f(h) \sim 1 - \xi (1-h)+D \Gamma(1-\gamma) (1-h)^{\gamma -1} \, \text{ as } h\to 1^-,
\end{equation}
where $\Gamma(\cdot)$ is the Gamma function. Using this asymptotic series, the integrands in both Eqs.~(\ref{23}) and (\ref{SVintegral}) have the leading order behaviour $\sim (1-h)^{2-\gamma}$ as $h\to 1^-$ at the critical value of $\xi=1$. Since this singularity is integrable, the resulting values of $d$ and $s$ are both finite at $\xi=1$, in contrast to the divergence seen in the case where $f''(1)$ is finite. The example of the solid curves in Fig.~\ref{figasympt} is for the offspring distribution where $q_\ell \propto \ell^{-\gamma}$ for all $\ell\ge 1$ (and $q_0=1-\sum_{\ell>0}q_\ell$), with power-law exponent $\gamma=2.5$. The finite limits of $d$ and $s$ as the branching number approaches 1 are evident.

\subsection{Novelty decay in a multiplicative process model} \label{novelty}

Multiplicative stochastic processes have been used in a number of papers to model popularity growth \cite{WuH07,Yasseri17}. In our notation, the assumption of the multiplicative model is that the total number of tweets by generation $n$ (i.e., the tree size $\widetilde{X}_n$) can be considered as proportional to the number of tweets that occurred in all previous generations ($\widetilde{X}_{n-1}$), multiplied by a random factor $W_n$ that is modulated by a novelty decay factor $r_n$:
\begin{equation}
\widetilde{X}_n = \left(1+r_n W_n\right) \widetilde{X}_{n-1}. \label{nov1}
\end{equation}
Here, the random variables $W_1, W_2, \ldots$ are assumed to be positive, independent, and identically distributed for each tree, while $r_n$ is a deterministic novelty decay factor that is common to all trees.

The novelty decay factor for this model can be obtained from Eq.~(\ref{nov1}) by rewriting it as
\begin{equation}
r_n W_n = \frac{\widetilde{X}_n - \widetilde{X}_{n-1}}{\widetilde{X}_{n-1}} \label{nov2}
\end{equation}
and taking expectations (i.e., averaging over all trees). The deterministic novelty decay factor $r_n$ is then proportional to the expectation of the right hand side of Eq.~(\ref{nov2}). Using the fact that the number of particles $\widetilde{Z}_n$ in generation $n$ can be related to the tree size by
\begin{equation}
\widetilde{Z}_n = \widetilde{X}_n - \widetilde{X}_{n-1},
\end{equation}
we therefore consider  the calculation of the quantity
\begin{equation}
\widetilde{r}_n = E\left(\frac{\widetilde{Z}_n}{\widetilde{X}_n - \widetilde{Z}_n}\right), \label{r1}
\end{equation}
which, up to a multiplicative constant, is the novelty decay factor in such models. (The multiplicative constant is often set, as in \cite{WuH07} for example, by normalizing the value of $r_1$).

Similar to Eq.~(\ref{13}), we consider here the joint distribution, at generation $n$, of tree size $\widetilde{X}_n$ and number of particles $\widetilde{Z}_n$, defining the two-variable pgf $\widetilde{K}_n(x,z)$ as
\begin{equation}
\widetilde{K}_n(x,z) = E\left( x^{\widetilde{X}_n} z^{\widetilde{Z}_n}\right) = \sum_{j, \ell = 0}^\infty \text{Prob}\left(\widetilde{X}_n=j\text{ and }\widetilde{Z}_n = \ell\right)x^j z^\ell.
\end{equation}
The iteration equation for $\widetilde{K}_n$ is, similar to Eqs.~(\ref{10}) and (\ref{4}),
\begin{equation}
\widetilde{K}_n(x,z) = x \widetilde{f}\left(K_{n-1}(x,z)\right), \label{r2}
\end{equation}
where $K_n(x,z)$ satisfies
\begin{equation}
K_n(x,z) = x f\left(K_{n-1}(x,z)\right),
\end{equation}
and $K_0(x,z) = x z$.

We observe that if we modify the second argument of $\widetilde{K}$ as follows
\begin{align}
\widetilde{K}_n\left(x,\frac{z}{x}\right) &=\sum_{j, \ell = 0}^\infty \text{Prob}\left(\widetilde{X}_n=j\text{ and }\widetilde{Z}_n = \ell\right)x^j \left(\frac{z}{x}\right)^\ell\\
&= \sum_{j, \ell = 0}^\infty \text{Prob}\left(\widetilde{X}_n=j\text{ and }\widetilde{Z}_n = \ell\right)x^{j-\ell} z^\ell  ,
\end{align}
then we can write, analogous to Eq.~(\ref{17}),
\begin{align}
\widetilde{r}_n & = \sum_{j,\ell=0}^\infty \text{Prob}\left(\widetilde{X}_n=j\text{ and }\widetilde{Z}_n = \ell\right) \frac{\ell}{j-\ell} \\
&= \int_0^1 \frac{1}{x} \left. \frac{\partial \widetilde{J}_n}{\partial z}\right|_{z=1} dx, \label{r55}
\end{align}
where $\widetilde{J}_n(x,z) = \widetilde{K}_n\left(x,\frac{z}{x}\right)$. The iteration equation for $\widetilde{J}_n(x,z)$ is obtained from Eq.~(\ref{r2}) as
\begin{equation}
\widetilde{J}_n(x,z)) = x \widetilde{f}\left(J_{n-1}(x,z)\right), \label{r3}
\end{equation}
where
\begin{equation}
J_n(x,z)= K_n\left(x,\frac{z}{x}\right) = x f \left(J_{n-1}(x,z)\right), \label{Jeqn}
\end{equation}
and $J_0(x,z)=z$.

To evaluate the integral in Eq.~(\ref{r55}), it is convenient to define the single-argument function
\begin{equation}
\widetilde{L}_n(x) = \frac{1}{x} \left. \frac{\partial \widetilde{J}_n}{\partial z}\right|_{z=1} = \frac{1}{x} \left. \frac{\partial \widetilde{K}_n\left(x,\frac{z}{x}\right)}{\partial z}\right|_{z=1}.
\end{equation}
Then we obtain from Eq.~(\ref{r3}) that $\widetilde{L}_n$ can be expressed as
\begin{equation}
\widetilde{L}_n(x) = x \widetilde{f}'\left(J_{n-1}(x,1)\right) L_{n-1}(x),\label{Ltildeeqn}
\end{equation}
where $L_n(x)$, defined as
\begin{equation}
L_n(x) = \left. \frac{1}{x} \frac{\partial {J}_n}{\partial z}\right|_{z=1},
\end{equation}
obeys the iteration equation
\begin{equation}
L_n(x) = x f'\left(J_{n-1}(x,1)\right) L_{n-1}(x), \label{Leqn}
\end{equation}
with $L_0(x)=1/x$.
Iterating Eqs.~(\ref{Ltildeeqn}), (\ref{Leqn}) and (\ref{Jeqn}) for values of $x$ that partition the interval $[0,1]$ enables us to calculate the integral
\begin{equation}
\widetilde{r}_n = \int_0^1 \widetilde{L}_n(x) dx \label{rtheory}
\end{equation}
using the trapezoidal rule.

Thus, we have shown how a subcritical branching process model can give rise to an apparent novelty decay factor, even though the offspring distribution does not change from generation to generation. The ``apparent'' nature of the decay in the novelty factor does not reflect any change in the likelihood of retweeting by a user who receives the tweet; rather it is the mechanism needed in the multiplicative process model of Eq.~(\ref{nov1}) to deal with the finite lifetimes of cascades. At each generation of the branching process fewer trees remain alive, and so the growth rate of the total number of tweets must decline with $n$, and in the multiplicative process this is mediated by the decay of the novelty factor $r_n$.

\begin{figure}
\centering
\epsfig{figure=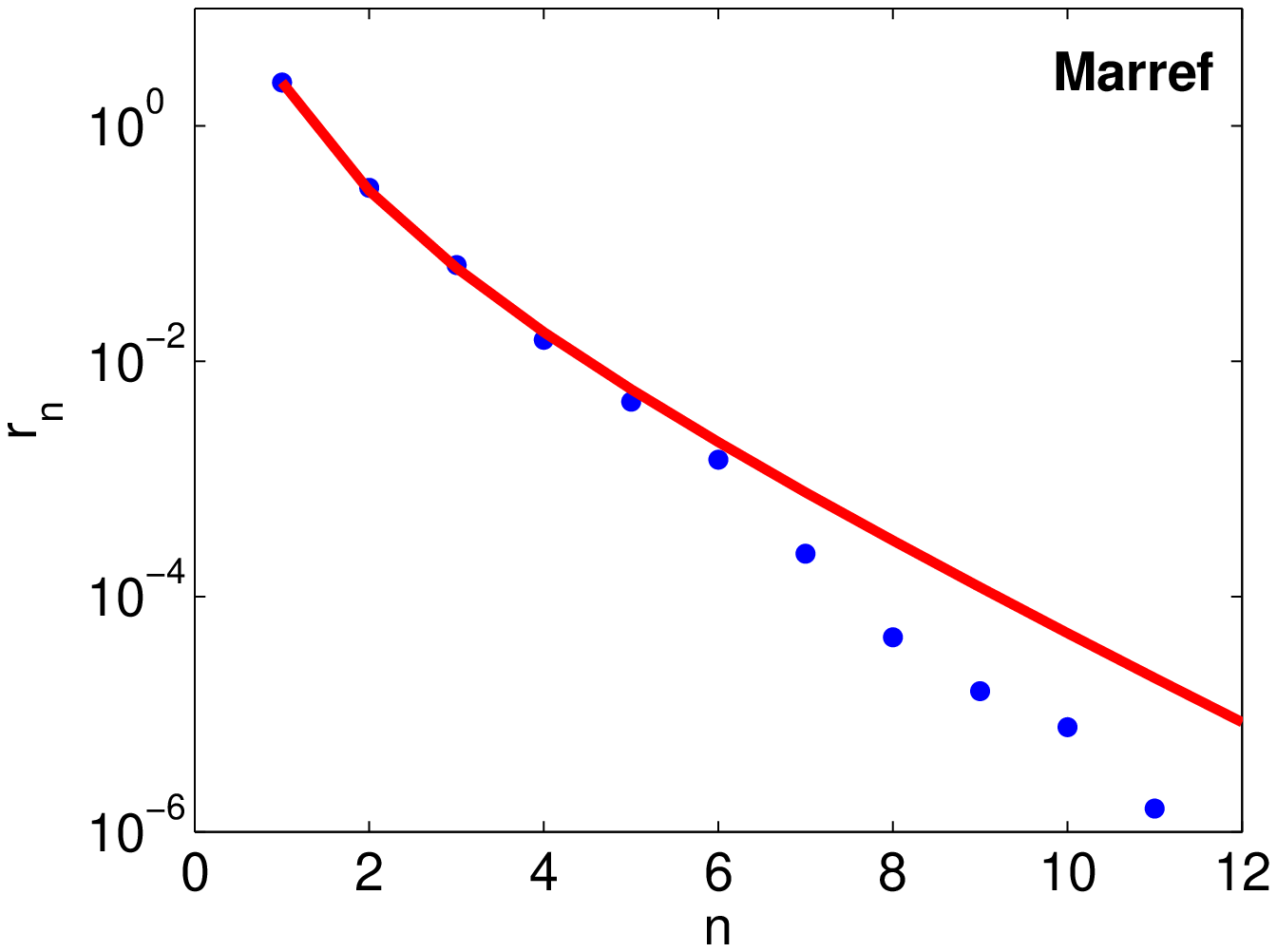,width=8.1 cm}
\epsfig{figure=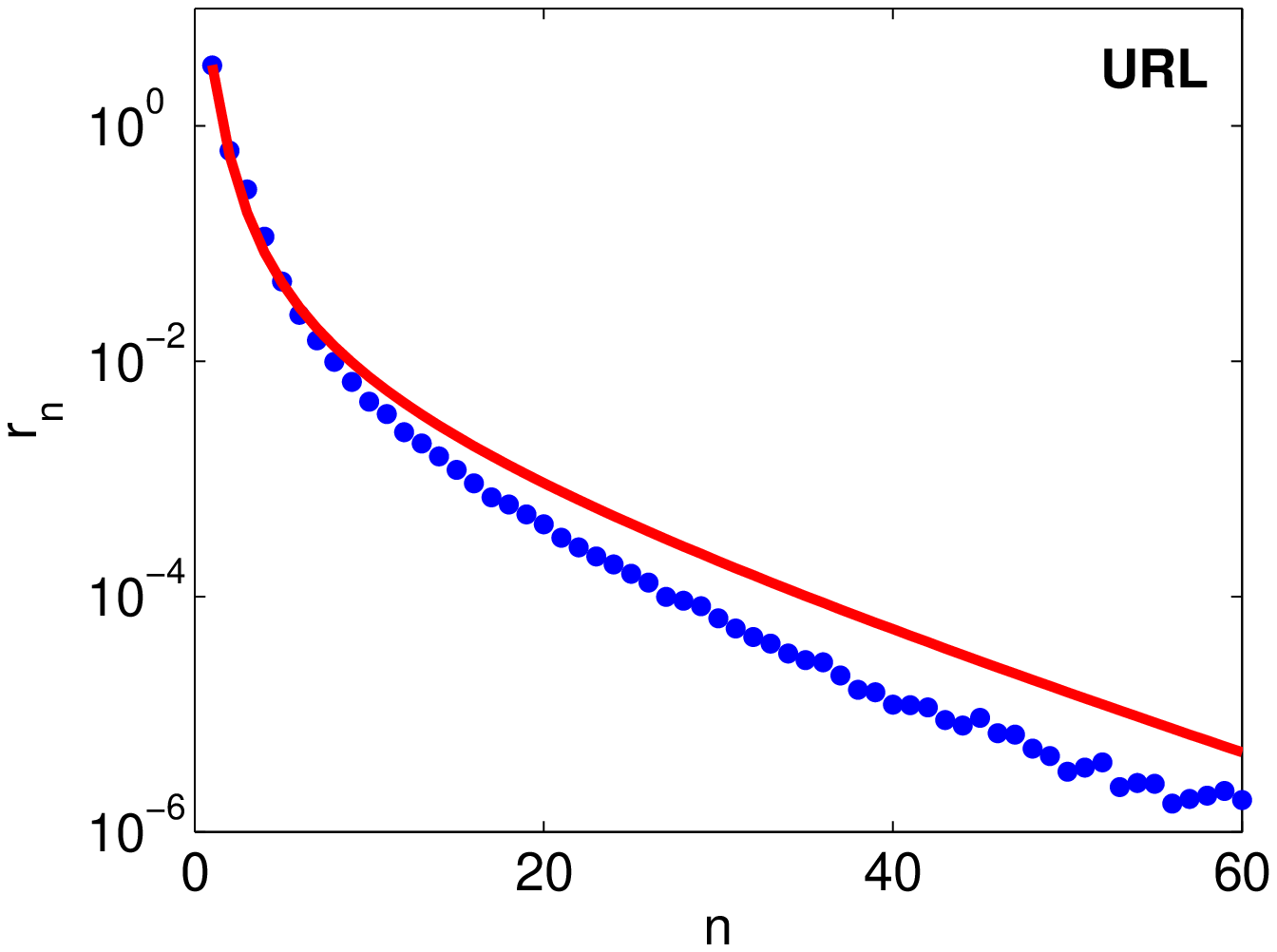,width=8.1 cm} 
\caption{Novelty function $\widetilde{r}_n$ in Marref (left) and URL (right) datasets. Blue symbols are empirical values using Eq.~(\ref{r1}); red lines are the predictions of the theoretical result (\ref{rtheory}), using the offspring distribution of Eq.~(\ref{3.15}). }\label{fig6}
\end{figure}

\section{Discussion} \label{sec5}

In Section~\ref{sec2} we demonstrated that two datasets from Twitter can be approximately described by branching processes, at least when we examine the discrete generation-by-generation structure. An examination of the details of a continuous-time branching process that could produce these structures is left for further work. In Section~\ref{sec3} we argued that the observed offspring distributions were better fitted by a model based on the assumption that Twitter users have limited attention---so those who follow many others are less likely to  notice and retweet any single message they receive---than by the more usual independent cascade model, with its assumption of equal transmission probability for each infection attempt.

Taking the fitted offspring distributions as inputs, in Section~\ref{sec4} we derived analytical and semi-analytical results using branching process theory. We began with well-established results on the distribution of cascade lifetimes and of cascade sizes, and then extended the arguments used to derive novel results for other measures. We derived integral formulas for the expected average tree depth (equation~(\ref{23})) and for the expected structural virality (equation~(\ref{SVintegral})) and showed that these provide a good match to the data. The integral formulas are also amenable to asymptotic analysis to understand the behaviour of the metrics as the branching number approaches the critical value. These results should assist in the fitting of transmission models to large-scale datasets, as was done (albeit using billions of numerical simulations rather than analytical methods) in Goel et al.~\cite{Goel15}. Finally, we derived a formula that enables the calculation of the apparent novelty factor, as would be used in a multiplicative stochastic model for the cascades under study. In the branching process model, information does not decrease in its transmission likelihood over generations, but the fact that the processes are subcritical means that the number of users who receive a cascading tweet decreases over time (Figure~\ref{fig2}). In a multiplicative model, the stochastic lifetimes of cascade trees must be imposed through the assumption of novelty decay, and our results in Sec.~\ref{novelty} show how the two modelling approaches can be directly compared. We believe that the insights of the branching process approach will help inform applications of the multiplicative model, while the formula linking the offspring distribution to the apparent novelty decay (equation~(\ref{rtheory})) will allow the application of branching process theory to datasets that previously were studied only via the multiplicative model.

Our study has, of course, several limitations. The nature of cascades on Twitter is that they are rather short-lived, so our observation of a stable offspring distribution might not generalize to cascades on other social media where the attention given to topics is longer-lived, and hence where novelty decay might be more likely. We have implicitly assumed that all cascade topics are equally attractive to the Twitter users and so the identification of cascade-specific ``fitnesses'' \cite{Yook20} has not been addressed here. As noted above, a study based on continuous-time branching processes could potentially extend our results to include age-dependent effects \cite{Iribarren11}, but we expect that the results presented here would remain valid in the long-time limit where all cascades have reached their final state.
In conclusion, we hope that the results and the methodology presented here will prove useful to researchers investigating those aspects of human behaviour that are mediated by online social networks.

\section*{Acknowledgements}
This work is partly supported by Science Foundation Ireland (grant numbers 16/IA/4470, 16/RC/3918, 12/RC/2289 P2 and  18/CRT/6049) with co-funding from the European Regional Development Fund (J.G.), by the James S.~McDonnell Foundation (P.F.) and by JSPS KAKENHI (grant number JP19K14618) (T.O.). We acknowledge the work of the authors of \cite{OSullivan17, Lerman12, Hodas14} in gathering the initial datasets and making them available for study.
%

\bibliographystyle{unsrt}
\bibliography{trees_refs}

\end{document}